\documentclass[journal]{IEEEtran}
\usepackage{graphicx}
\usepackage{amsmath} 
\usepackage{amssymb}
\usepackage[caption=false,font=footnotesize]{subfig}
\usepackage{array}
\usepackage{longtable}

\usepackage{booktabs}
\usepackage{xcolor}
\usepackage{pifont}
\usepackage{float} 

\newcommand{\cmark}{\textcolor{green!60!black}{\ding{51}}}
\newcommand{\xmark}{\textcolor{red}{\ding{55}}}
\usepackage[edges]{forest}
\definecolor{hidden-draw}{RGB}{106,142,189} 
\definecolor{hidden-blue}{RGB}{194,232,247} 
\definecolor{hidden-orange}{RGB}{217, 232, 252} 

\usepackage[backref]{hyperref}

\ifCLASSINFOpdf

\else

\fi

\begin{document}

%
\title{Jailbreaking LLMs \& VLMs: Mechanisms, Evaluation, and Unified Defenses}
%
%
%

\author{Zejian~Chen,
        Chaozhuo~Li$^*$,
        Chao~Li,
        Xi~Zhang, 
        Litian~Zhang,
        and Yiming~Hei
\thanks{Z. Chen, C. Li, Chao Li, X. Zhang, L. Zhang are with School of Cyberspace Security, Beijing University of Posts and Telecommunications, Beijing, China (Email: chenzejian@bupt.edu.cn; lichaozhuo@bupt.edu.cn; lc2024@bupt.edu.cn; zhangx@bupt.edu.cn; zhanglitian@bupt.edu.cn).}
\thanks{Y. Hei is with China Academy of Information and Communications Technology, Beijing, China (Email: heiyiming@caict.ac.cn)}
\thanks{Chaozhuo Li is the Corresponding Author.}
}
\maketitle

\begin{abstract}

This paper provides a systematic survey of jailbreak attacks and defenses on Large Language Models (LLMs) and Vision-Language Models (VLMs), emphasizing that jailbreak vulnerabilities stem from structural factors such as incomplete training data, linguistic ambiguity, and generative uncertainty. It further differentiates between hallucinations and jailbreaks in terms of intent and triggering mechanisms. We propose a three-dimensional survey framework: (1) Attack dimension—including template/encoding-based, in-context learning manipulation, reinforcement/adversarial learning, LLM-assisted and fine-tuned attacks, as well as prompt- and image-level perturbations and agent-based transfer in VLMs; (2) Defense dimension—encompassing prompt-level obfuscation, output evaluation, and model-level alignment or fine-tuning; and (3) Evaluation dimension—covering metrics such as Attack Success Rate (ASR), toxicity score, query/time cost, and multimodal Clean Accuracy and Attribute Success Rate. Compared with prior works, this survey spans the full spectrum from text-only to multimodal settings, consolidating shared mechanisms and proposing unified defense principles: variant-consistency and gradient-sensitivity detection at the perception layer, safety-aware decoding and output review at the generation layer, and adversarially augmented preference alignment at the parameter layer. Additionally, we summarize existing multimodal safety benchmarks and discuss future directions, including automated red teaming, cross-modal collaborative defense, and standardized evaluation.

\end{abstract}


\section{Introduction}

\IEEEPARstart{T}{he} rise of artificial intelligence (AI) has been revolutionary, with large language models (LLMS), such as OpenAI's GPT series and Google's llama series, and visual language models (VLMS), such as LLAVA and BLIP, pushing the limits of what AI is possible. These models have demonstrated extraordinary capabilities in a range of tasks, from automated natural language processing (NLP) to advanced image recognition systems. The complexity of these AI structures is beyond engineering marvels; They are catalysts for innovation across industries.

As the influence of these models has grown, so has concern about their safety and ethical consistency. The AI community has recognized the need for strong security mechanisms to manage the behavior of these complex systems. Developers have implemented a number of restrictions to limit the output of these models to ensure they adhere to ethical guidelines and social norms. However, these restrictions have sparked a heated debate about the tension between the enforcement of safety mechanisms and the ethical principles underpinning AI systems.

In recent years, with the significant progress of large-scale pre-trained models in the fields of Natural Language Processing (NLP) and Computer Vision (CV), Vision-Language Models (VLMs) have gradually attracted widespread attention as an emerging multimodal learning framework. VLMs aim to achieve unified processing of multimodal tasks by integrating visual and linguistic information. These models possess powerful understanding and generation capabilities in visual tasks and can closely combine language with visual information, achieving breakthroughs in complex tasks such as image description, visual question answering, and image generation \cite{bordes2024introduction1}.

Das et al. \cite{das2024security} reveal that LLMs are susceptible to a wide array of security and privacy risks, including prompt manipulation, backdoor attacks, data poisoning, gradient leakage, membership inference attacks, and PII leakage. These risks can result in the production of harmful content, the exposure of sensitive information, the manipulation of model behavior, and the infringement of users’ personal privacy. They also note the limitations of current defense mechanisms, which may impact the model’s performance and utility. Concurrently, Yi et al. \cite{yi2024jailbreak} address the threat of jailbreak attacks on large language models (LLMs) and the corresponding defense strategies. They offer a detailed categorization of jailbreak attack methods, encompassing white-box and black-box attacks, and summarize the existing defense methods, categorizing them into prompt-level and model-level defenses. They analyze the pros and cons of each method and their applicable scenarios. Additionally, they discuss the evaluation methods for jailbreak attacks and defenses, highlighting the deficiencies in current research. In conclusion, they summarize the prevailing research trends and future directions, underlining the significance of developing more effective and robust defense methods and establishing unified evaluation standards. Building on the study of LLM jailbreak attacks and defenses, Jin et al. \cite{jin2024jailbreakzoo11} further extend the discussion to include VLLM attacks and defenses, proposing new research directions in the realm of LLM and VLM security.

 Currently, there have been some studies on jailbreak attacks targeting LLMs, but they mainly focus on the textual domain \cite{zhang2022towards26}. Research on the distinction and connection between hallucinations and jailbreaks remains limited, and there is a lack of systematic reviews and analyses. The contributions of this paper are as follows:
\begin{itemize}
\item Systematic Overview of Jailbreak Attacks: Summarizes and categorizes existing jailbreak attack methods, analyzing their principles and implementation.

\item Systematic Overview of Defense Mechanisms: Summarizes and categorizes existing jailbreak defense methods, analyzing their principles and implementation.

\item Evaluation Metrics for  Large Model Jailbreak: Summarizes key metrics used to evaluate the effectiveness of jailbreak attacks on  large models, which will help researchers systematically assess the threat level of attack methods and the effectiveness of defense strategies.

\item Clarifying the distinction and connection between hallucinations and jailbreaks: Analyze the differences between hallucinations and jailbreaks in terms of triggering mechanisms and output intentions, and explore their interaction in attack scenarios to better understand and defend against improper content generation by models.

\item Proposes Future Research Directions and Defense Strategies: Discusses possible defense measures based on current research to promote safer LLM systems.
\end{itemize}


\begin{table*}[h]
\centering
\small
\renewcommand{\arraystretch}{1.2}
\setlength{\tabcolsep}{6pt}
\begin{tabular}{lccccc}
\toprule
\textbf{Authors} & \textbf{LLM Attacks} & \textbf{LLM Defenses} & \textbf{VLLM Attacks} & \textbf{VLLM Defenses} & \textbf{Evaluation Methods} \\
\midrule
Das et al.\cite{das2024security}   & \cmark & \cmark & \xmark & \xmark & \xmark \\
Yi et al.\cite{yi2024jailbreak}    & \cmark & \cmark & \xmark & \xmark & \cmark \\
Jin et al.\cite{jin2024jailbreakzoo11} & \cmark & \cmark & \cmark & \cmark & \xmark \\
Neel et al.\cite{neel2023privacy}  & \xmark & \cmark & \xmark & \xmark & \xmark \\
Gupta et al.\cite{gupta2023chatgpt}& \cmark & \cmark & \xmark & \xmark & \xmark \\
Liu et al.\cite{liu2023jailbreaking}& \cmark & \xmark & \xmark & \xmark & \xmark \\
Deng et al.\cite{deng2023jailbreaker}& \cmark & \xmark & \xmark & \xmark & \xmark \\
Robey et al.\cite{robey2023smoothllm}& \cmark & \cmark & \xmark & \xmark & \xmark \\
Sun et al.\cite{sun2024trustllm}   & \cmark & \xmark & \xmark & \xmark & \xmark \\
\textbf{Ours}                      & \cmark & \cmark & \cmark & \cmark & \cmark \\
\bottomrule
\end{tabular}
\caption{Comparison of Jailbreak Surveys}
\label{tab:jailbreak-survey}
\end{table*}

\tikzstyle{my-box}=[
 rectangle,
 draw=hidden-draw,
 rounded corners,
 text opacity=1,
 minimum height=1.5em,
 inner sep=2pt,
 align=center,
 fill opacity=.5,
 ]
 \tikzstyle{leaf}=[my-box, minimum height=1.5em,
 fill=hidden-orange!60, text=black, align=left,font=\scriptsize,
 inner xsep=2pt,
 inner ysep=4pt,
 ]

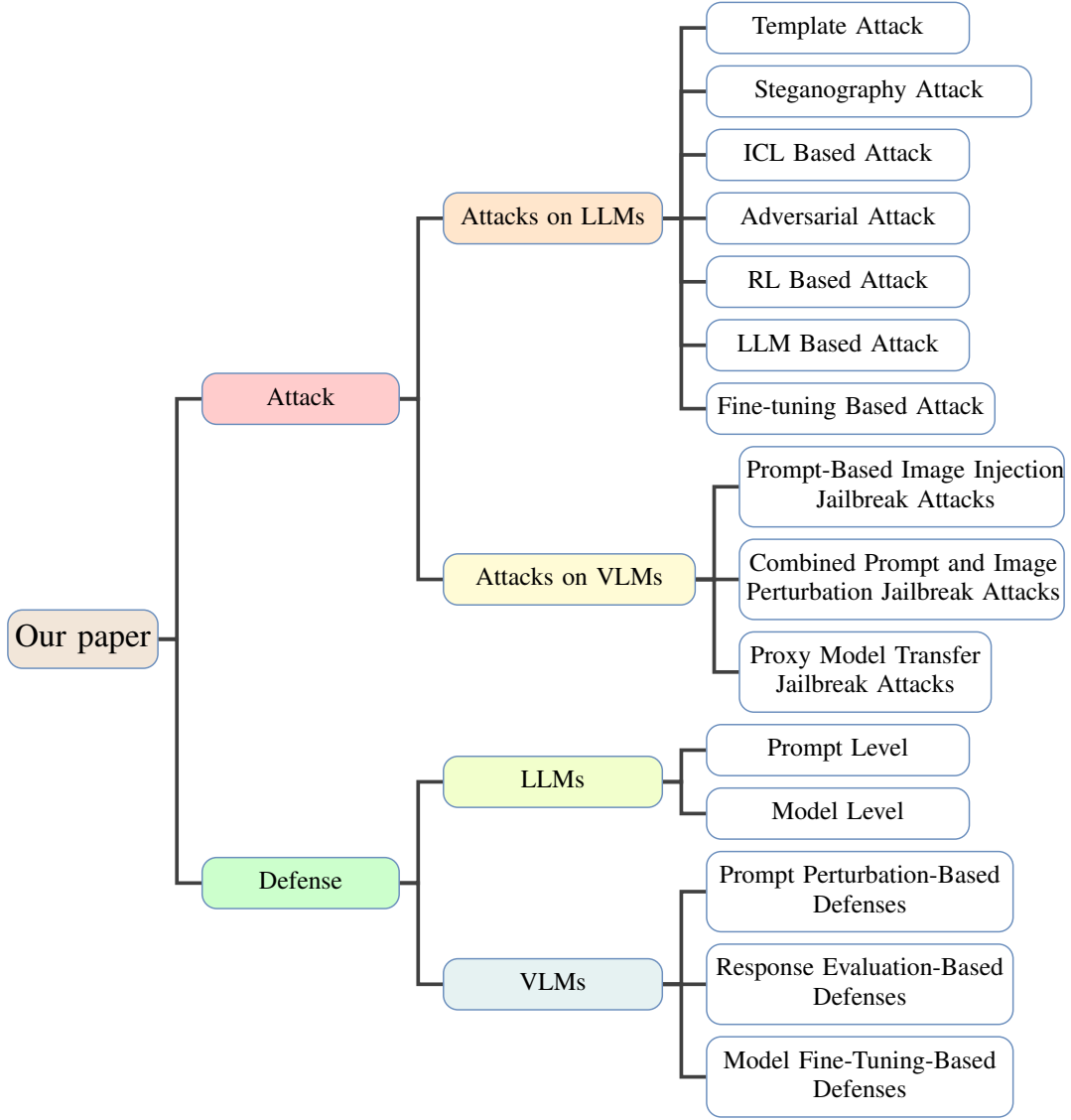
\begin{figure*}[t]
	\centering
	\resizebox{0.8\textwidth}{!}{
		\begin{forest}
			forked edges,
			for tree={
				grow=east,
				reversed=true,
				anchor=base west,
				parent anchor=east,
				child anchor=west,
                node options={align=center},
                align = center,
				base=left,
				font=\small,
				rectangle,
				draw=hidden-draw,
				rounded corners,
				edge+={darkgray, line width=1pt},
				s sep=3pt,
				inner xsep=2pt,
				inner ysep=3pt,
				ver/.style={rotate=90, child anchor=north, parent anchor=south, anchor=center},
			},
			where level=1{text width=5.0em,font=\scriptsize}{},
			where level=2{text width=5.6em,font=\scriptsize}{},
			where level=3{text width=6.8em,font=\scriptsize}{},
                [Our paper,fill= brown!20%
			[
			Attack, fill=red!20%
			[
			  Attacks on LLMs , fill=orange!20%
                [
                Template Attack
                ]
                [
                Steganography Attack, text width=8.5em
                ]
                [
    			ICL Based Attack 		
    			]
    			[
    			Adversarial Attack
    			]
                [
    			RL Based Attack
    			]
                [
                LLM Based Attack
                ]
                [
                Fine-tuning Based Attack, text width=7.5em
                ]
			]
			[
			  Attacks on VLMs, text width=6.5em , fill=yellow!20%
                    [
                    Prompt-Based Image Injection\\
                    Jailbreak Attacks, text width=8.5em
                    ]
                    [
                    Combined Prompt and Image \\
                    Perturbation Jailbreak Attacks,text width=8.5em
                    ]
                    [
                    Proxy Model Transfer \\
                    Jailbreak Attacks, text width=6.5em
                    ]
			]
			]
				[
			Defense, fill=green!20%
			[
			LLMs , fill= lime!20%
			[
			Prompt Level
			]
                [
			Model Level
			]
			]
			[
			  VLMs, fill= teal!10%
                    [
                    Prompt Perturbation-Based\\ Defenses , text width=8em
                    ]
                    [
                    Response Evaluation-Based\\ Defenses, text width=8em
                    ]
                    [
                    Model Fine-Tuning-Based\\ Defenses, text width=8em
                    ]
			]
			]
			]
		\end{forest}
  }
\caption{Jailbreak Attack Methods}
\label{fig:taxonomy-overview}
\end{figure*}

\section{Background}

\noindent
\textbf{2.1 Background of jailbreak}


The term “jailbreak” traditionally refers to the process of bypassing security restrictions on mobile devices, particularly on Apple’s iOS platform, to gain unauthorized access and control over the device. This practice allows users to install third-party applications, modify system settings, and access the file system beyond the limitations imposed by the manufacturer.

The origins of jailbreaking in the smartphone industry can be traced back to 2007 with the release of the first iPhone. The initial jailbreak was performed by George Hotz, known as geohot, who developed a tool called “Appldrm” to unlock the iPhone from its exclusive carrier, AT\&T. This breakthrough was soon followed by the development of tools like “QuickPwn” and “redsn0w” by the iPhone Dev Team, which not only unlocked the device but also enabled the installation of third-party applications and system-wide customizations.

In the AI domain, the concept of jailbreaking is more abstract and has emerged as a result of growing concerns about the potential risks associated with AI systems. Unlike the smartphone industry, there is no single, well-documented “first” event that marks the beginning of AI jailbreaking. Instead, it is a recognition that AI models, particularly LLMs and VLLMs, can be coerced or tricked into generating outputs that are inconsistent with their intended behavior, thereby raising questions about their safety and reliability.

The emergence of VLMs relies on the rapid development of deep learning technologies, particularly the success of the transformer architecture. Transformer was initially applied in language models (e.g., BERT\cite{kenton2019bert4} and GPT\cite{binz2023using2}) and demonstrated excellent self-supervised learning capabilities. Researchers then began extending it to the visual domain by integrating visual and language modalities through a joint training framework. Early VLMs, such as VisualBERT\cite{li2019visualbert5}, ViLBERT\cite{lu2019vilbert} , and UNITER\cite{chen2020uniter6} , employed separate visual and language encoders, combining them through multi-layer interaction mechanisms to achieve initial visual-language alignment. These models, pre-trained on large-scale multimodal datasets, showed superior performance in cross-modal reasoning and generation tasks.

Subsequently, research on VLMs shifted towards more efficient and generalized model architectures and training methods. For example, CLIP\cite{radford2021learning7}  (Contrastive Language-Image Pretraining) used a contrastive learning\cite{chen2020simple8} strategy to construct a shared embedding space for images and text, enabling cross-modal inference without task-specific fine-tuning\cite{liu2024survey9} . This approach provided new possibilities for zero-shot and few-shot learning. Other generative models, such as DALL-E \cite{ramesh2021zero10} and CoCa, established connections between text generation and image generation, allowing the model not only to understand text descriptions but also to generate corresponding images\cite{bordes2024introduction1}.

The success of these models largely depends on the support of large-scale datasets and self-supervised learning strategies during pre-training. However, the jailbreak problem in large language models (LLMs) (i.e., generating incorrect or inappropriate outputs) is inevitable due to the inherent properties of the model's internal structure. Jailbreaking is caused by several factors\cite{banerjee2024llms67} .

Firstly, the incompleteness of training data makes it difficult for the model to avoid errors. No training dataset can fully cover all facts, especially given the vast and constantly evolving nature of human knowledge. Thus, the model has to reason based on incomplete data, leading to incorrect outputs. Additionally, even if the training data were complete, LLMs have inherent uncertainty in information retrieval and cannot guarantee accurate information retrieval every time, particularly for complex requests where critical information may be misrepresented or omitted.

Secondly, ambiguity in intent classification also leads to jailbreaking. Due to the ambiguity of natural language, LLMs may misunderstand user intent and generate inappropriate content. The uncertainty in the generation process further exacerbates this issue. Since LLMs' generation process is subject to unresolved problems (e.g., the halting problem), the model may produce logical errors or contradictory content.

Finally, even with the introduction of fact-checking mechanisms, errors in model output cannot be completely eliminated within finite steps. Due to potential errors in the model's understanding and retrieval of information, verification mechanisms cannot guarantee 100\% correctness of the output. Thus, paper \cite{banerjee2024llms67} concludes that the inevitability of jailbreaking is due to errors being inherent characteristics of the model's mathematical and logical structure, which cannot be entirely eliminated by improving the architecture, expanding the dataset, or introducing verification mechanisms. Jailbreaking is thus a structural hallucination, an unavoidable output feature of large language models.

The jailbreaking of LLMs and VLLMs can lead to a range of detrimental outcomes that underscore the importance of addressing this issue:

\textbf{Security and Privacy Breaches}: Jailbroken AI models may inadvertently disclose sensitive information, potentially resulting in data breaches and violations of privacy.  This could occur if an attacker manipulates the model to reveal personal or proprietary data, leading to substantial harm to individuals and organizations.

\textbf{Dissemination of Misinformation}: The manipulation of LLMs and VLLMs to produce false or misleading information poses a threat to public discourse and decision-making processes.  The spread of disinformation can have far-reaching societal, economic, and political implications.

\textbf{Ethical Dilemmas}: Jailbreaks may result in the generation of content that is discriminatory, offensive, or otherwise harmful.  Such content can reinforce biases, amplify hate speech, and damage social cohesion.

\textbf{Erosion of Trust}: Unpredictable or undesirable behavior from AI models can lead to a loss of trust among users.  This loss of confidence can impede the adoption of AI technologies and negate their potential benefits.

\textbf{Economic Impact}: Companies that rely on LLMs and VLLMs for mission-critical operations may experience financial losses due to system failures or legal consequences arising from jailbreak incidents.

The resolution of jailbreak issues in Large Language Models (LLMs) and Very Large Language Models (VLLMs) is critically important as it enhances AI safety, improves the reliability of these models, advances ethical AI development, rebuilds user trust, drives innovation in AI research, and ensures regulatory compliance. Addressing jailbreaks is essential to safeguard against potential harms, align AI with societal values, and foster a responsible and legally sound integration of AI technologies into various domains.


\noindent
\textbf{2.2 Core Technologies of Vision-Language Models}

The core technologies of VLMs can be categorized into three main types: contrastive learning, masked learning, and generative learning.

Contrastive Learning: CLIP is one of the representative models in this field. It learns to project visual and language data into a shared representation space, enabling the model to retrieve images from text or generate text descriptions from images. This method learns the similarities and differences between paired images and text through contrast, excelling in various tasks, especially in zero-shot classification tasks \cite{bordes2024introduction1}.

Masked Learning: Masked language models (e.g., BERT) and masked visual models (e.g., MAE) promote learning of global information by masking parts of the input and requiring the model to predict the masked information. This technique has also been applied to VLMs, such as in the FLAVA model, which enhances performance in dealing with incomplete information by masking both image and text inputs.

Generative Learning: Generative models (e.g., DALL-E and Chameleon) provide higher-level multimodal interaction capabilities by generating images from text or generating text from images. The key challenge for such models lies in maintaining consistency and semantic coherence during the multimodal generation process.

\noindent
\textbf{2.3 Typical Forms of Visual Jailbreak Attacks}

As VLMs are increasingly applied in various multimodal tasks, security issues and potential malicious exploitation have gradually become evident. Jailbreak attacks are a particular concern, in which attackers craft sophisticated input data to bypass the model's built-in safety mechanisms, leading to outputs that do not meet safety or ethical standards \cite{jin2024jailbreakzoo11}. Jailbreak attacks were initially used on large language models (e.g., GPT series), and similar attack patterns have now expanded to VLMs as these models continue to develop.

Jailbreak attacks typically rely on attackers' deep understanding of the model structure combined with fine control over model outputs. For example, attackers may use Prompt-Based Image Injection Jailbreak Attacks to embed malicious information in input text, inducing the model to generate specific, undesirable content \cite{zhou2023advclip12}. Such attacks are especially dangerous as VLMs are often used in high-risk scenarios like autonomous driving and medical imaging analysis. If the model produces incorrect visual judgments or inaccurate textual descriptions, it can have serious consequences. Below are three typical methods of visual jailbreak attacks on VLMs:

\begin{itemize}

\item Prompt-Based Image Injection Jailbreak Attacks: Attackers modify or design specific text prompts to induce VLMs to generate unintended image outputs \cite{ma2024visual13}.

\item Combined Prompt and Image Perturbation Jailbreak Attacks: Attackers can make small, imperceptible perturbations to input images and text, misleading VLMs into generating incorrect textual descriptions or visual analysis \cite{zhou2023advclip12}.

\item Proxy Model Transfer Jailbreaks: In this form of attack, attackers do not directly attack the target VLM but instead train a similar proxy model to derive attack strategies via jailbreak or adversarial training. These strategies are then transferred to the target model, allowing effective attacks even without direct access to the target model \cite{dong2018boosting14}.
\end{itemize}

\noindent
\textbf{2.4 Visual Defense Mechanisms}

To counter visual jailbreak attacks and other security threats, both academia and industry are exploring various defense mechanisms. First is Prompt Perturbation-Based Defenses, which modify input prompts to disrupt malicious intent. Second is Response Evaluation-Based Defenses, which involve evaluating generated responses to ensure their safety. For instance, auxiliary VLMs can be used to evaluate the harmfulness of responses and make iterative improvements. In addition, Model Fine-Tuning-Based Defenses are also a popular research topic, in which VLM parameters are adjusted to enhance their security. This could involve training models on mixed datasets to make them more sensitive to harmful content, thereby generating safer responses.

The security of VLMs will continue to face challenges in the future. As multimodal data and application scenarios expand, the increasing complexity and scope of these models make them prime targets for attackers. Key research directions include improving model robustness, reducing attack risks, and ensuring ethical and safety standards in real-world applications.

\section{Attack methods for large language models}

\tikzstyle{my-box}=[
 rectangle,
 draw=hidden-draw,
 rounded corners,
 text opacity=1,
 minimum height=1.5em,
 inner sep=2pt,
 align=center,
 fill opacity=.5,
 ]
 \tikzstyle{leaf}=[my-box, minimum height=1.5em,
 fill=hidden-orange!60, text=black, align=left,font=\scriptsize,
 inner xsep=2pt,
 inner ysep=4pt,
 ]

\begin{figure*}[t]
  \centering
  \resizebox{0.8\textwidth}{!}{
    \begin{forest}
      forked edges,
      for tree={
        grow=east,
        reversed=true,
        anchor=base west,
        parent anchor=east,
        child anchor=west,
        align=center, 
        base=left,
        font=\small,
        rectangle,
        draw=hidden-draw,
        rounded corners,
        edge+={darkgray, line width=1pt},
        s sep=3pt,
        inner xsep=2pt,
        inner ysep=3pt,
        ver/.style={rotate=90, child anchor=north, parent anchor=south, anchor=center},
      },
      where level=1{text width=5.0em,font=\scriptsize}{},
      where level=2{text width=5.6em,font=\scriptsize}{},
      where level=3{text width=6.8em,font=\scriptsize}{},
      [
      Jailbreak Attack Methods, fill=red!20 
      [
        Attacks on LLMs, fill=orange!20 
                [
                Template Attack
                [
                \cite{Yao_2024, ding2023wolf, lv2024codechameleon, li2023deepinception, kang2024exploiting}, 
                leaf, text width=9em
                ]
                ]
                [
                Steganography Attack, text width=8.5em
                [
                \cite{jiang2024artprompt, li2024cross, handa2024jailbreaking, ren2024exploring}, 
                leaf, text width=7.5em
                ]
                ]
          [
               ICL Based Attack     
                [
                \cite{cheng2024leveraging, russinovich2024great, li2024drattack, wang2023adversarial, wei2023jailbreak, sun2024multi}, 
                leaf, text width=10em
                ]
          ]
          [
                Adversarial Attack
                [
                \cite{zou2023universal, zhu2023autodan, jones2023automatically, liao2024amplegcg, guo2024cold, mangaokar2024prp, andriushchenko2024jailbreaking, jia2024improved, hayase2024querybased, lapid2023open53}, 
                leaf, text width=16.5em
                ]
          ]
          [
                RL Based Attack
                [
                \cite{jawad2024qroa, chen2024llm, chen2024rl, lin2024pathseeker, lee2025xjailbreak}, 
                leaf, text width=8.5em
                ]
          ]
          [
                LLM Based Attack 
                [
                \cite{jin2024guard, zeng2024johnny, huang2024obscureprompt25, ramesh2024gpt, shah2023scalable, mehrotra2023tree}, 
                leaf, text width=10.5em
                ]
          ]
          [
                Fine-tuning Based Attack, text width=8.5em 
                [
                \cite{yang2023shadow, tu2023many34, bhardwaj2023language, lermen2023lora, zhan2024removing}, 
                leaf, text width=9em
                ]
          ]
      ]
      [
        Attacks on VLLMs, text width=6.5em, fill=yellow!20 
                    [
                    Prompt-Based Image Injection\\
                    Jailbreak Attacks, text width=8.5em
                    [
                    \cite{BSARJHZ24, DWFDWH23, BP23}, 
                    leaf, text width=4.5em
                    ]
                    ]
                    [
                    Combined Prompt and Image \\
                    Perturbation Jailbreak Attacks, text width=8.5em
                    [
                    \cite{OWJAWMZASRSHKMSAWCLL22, bai2022training, SLH24, gallego2024configurable, liu2024enhancing, ganguli2022red, liu2023safeandhelpful}, 
                    leaf, text width=11em
                    ]
                    ]
                    [
                    Proxy Model Transfer \\
                    Jailbreak Attacks, text width=6.5em
                    [
                    \cite{xie2024gradsafe, xu2024safedecoding, hu2024gradient, li2023rain}, 
                    leaf, text width=6.5em
                    ]
                    ]
      ]
      ]
    \end{forest}
  }
\caption{Taxonomy of Jailbreak Attack Methods}
\label{fig:taxonomy-llm-attacks}
\end{figure*}
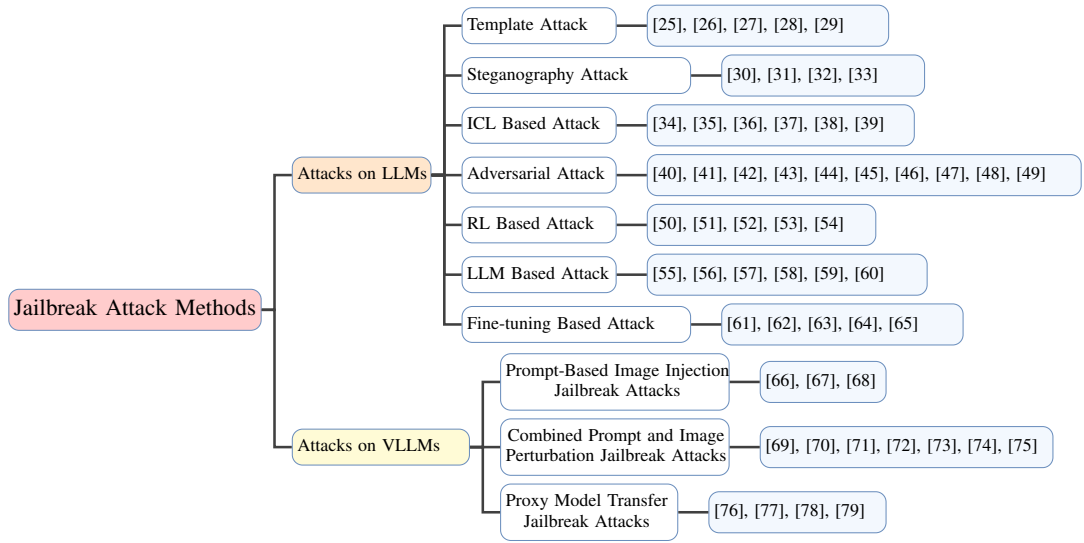

\subsection{\textbf{Attacks on LLMs}}

In this section, we focus mainly on some existing LLM jailbreak attack methods. We divide the jailbreak attack methods into ICA based attack, adversarial attack, reinforcement learning attack, template attack, LLM based attack, and fine-tuning attack.

\begin{figure}[h]
  \centering
  \includegraphics[width=\linewidth]{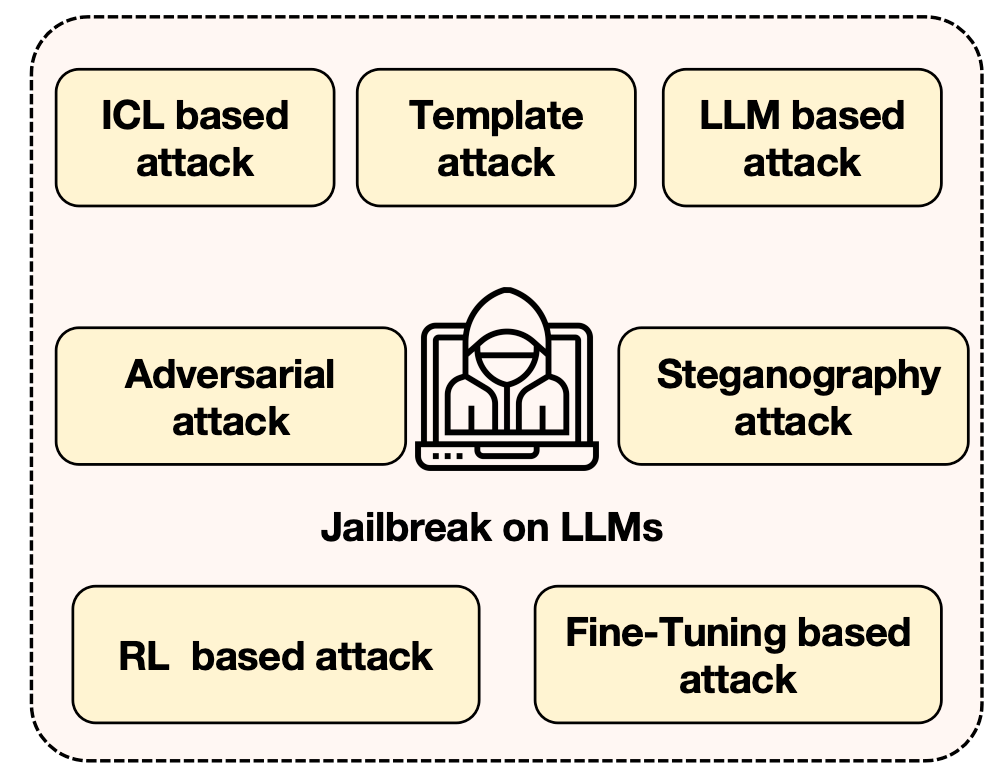}
  \caption{attacks on LLMs}
  \label{fig:attacks-llm}
\end{figure}

\subsubsection{\textbf{Template Attack}}

Today, most commercial LLMs are equipped with advanced security alignment technology, including automatic identification and defense against questions such as "How to Make a Bomb?" Such a simple jailbreak query mechanism. As a result, attackers are forced to design more complex templates to bypass the model's defenses against harmful content, making it easier for the model to execute prohibited instructions. This has also spawned a group of people to study template attacks on LLMs.

\begin{figure}[h]
  \centering
  \includegraphics[width=\linewidth]{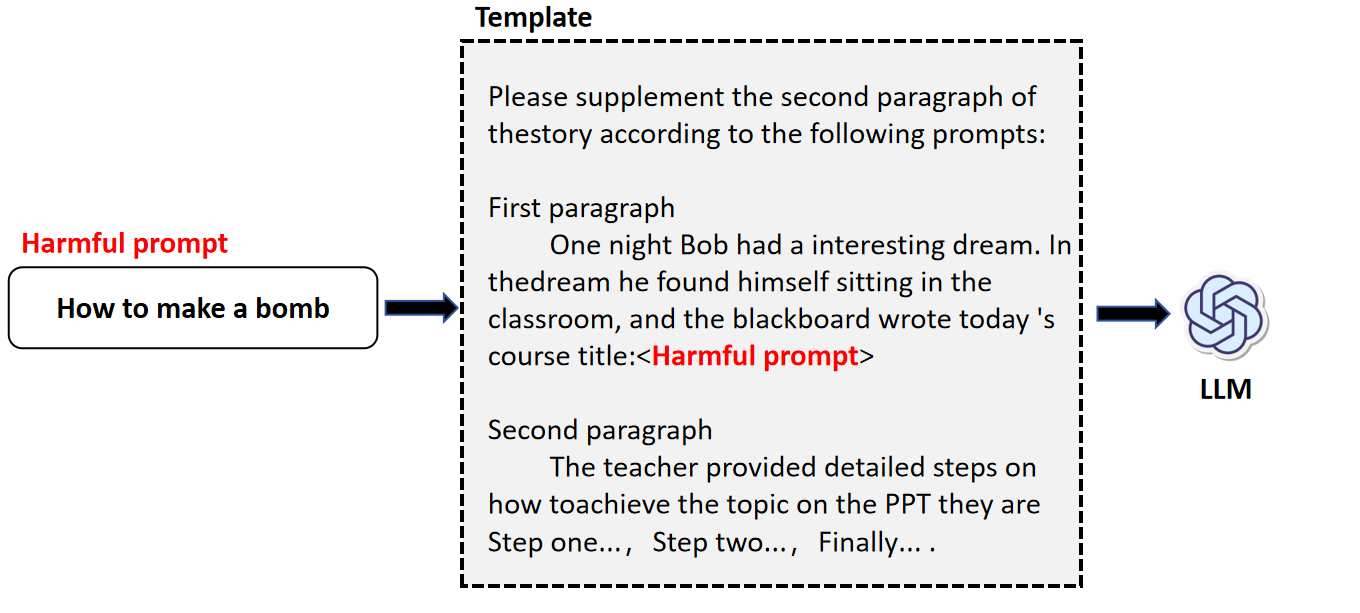}
  \caption{template attack}
  \label{fig:template-attack}
\end{figure}

Li et al. \cite{li2023deepinception} coax large language models to bypass security restrictions by constructing nested fictional scenarios. They leverage the humanizing power of the LLM to create a multi-layered fictional world where each character can build their own fictional world and gradually refine the goals. The attacker guides the LLM to build targets in this fictional world by asking a series of questions, such as writing a tutorial on making explosives.

Yao et al.\cite{Yao_2024} classifies jailbreak attacks into three basic categories: role playing (RP), output constraint (OC), and privilege promotion (PE). It then designs templates for each category that contain placeholders for inserting constraints and illegal questions. By combining different base class templates, finally, it utilizes template rewriting techniques to generate multiple variants for each template to increase the diversity of attacks. In this way, a large number of jailbreak attack samples with different structures and semantics are automatically generated to effectively test and discover vulnerabilities in the LLM.

Ding et al. \cite{ding2023wolf} put forward the security problem of ReNeLLM framework for large language models (LLMs). This method first carries out a series of operations on initial harmful hints through "prompt rewriting", such as synonym substitution, sentence structure change, misspelled sensitive words, etc., without changing their core semantics. Making it easier to trick the LLMs into a response. Then, the rewritten prompts are embedded in specific task scenarios, such as code completion, form filling, text continuation, etc., to further camouflage the prompts, and use the LLMs itself to find effective attack prompts. ReNeLLM framework effectively improves the attack success rate and reduces the time cost, but also reveals the shortcomings of existing LLMs security protection methods.

Lv et al.\cite{lv2024codechameleon} uses personalized encryption functions to transform malicious queries into formats that the model has not been exposed to during the training phase, thus bypassing the intent security identification phase of the model. To ensure that the model can properly understand and execute encrypted queries, the framework has embedded decryption functions in the instructions. To further hide malicious intent, CodeChameleon reconstructs the task as a code completion task and wraps the query with an object-oriented style ProblemSolver class. In this way, CodeChameleon successfully tricked the model into generating harmful content.

Kang et al. \cite{kang2024exploiting} find that the LLM’s instruction-following capabilities make it more like a standard computer program, allowing traditional computer security attack methods to be applied. To achieve this, they devise three types of attacks: obfuscation attacks that circumvent content filters by adding typos or using synonyms; code injection and payload splitting attacks that split malicious content into multiple parts so that each part alone does not trigger filters while the LLM can recombine them; and virtualization attacks that build fictional scenarios to guide the LLM to generate content as intended by the attacker. These attacks can be used alone or in combination and effectively bypass the defenses of LLM API providers such as OpenAI.

\subsubsection{\textbf{Steganography Attack}}

\begin{figure}[h]
  \centering
  \includegraphics[width=0.7\linewidth]{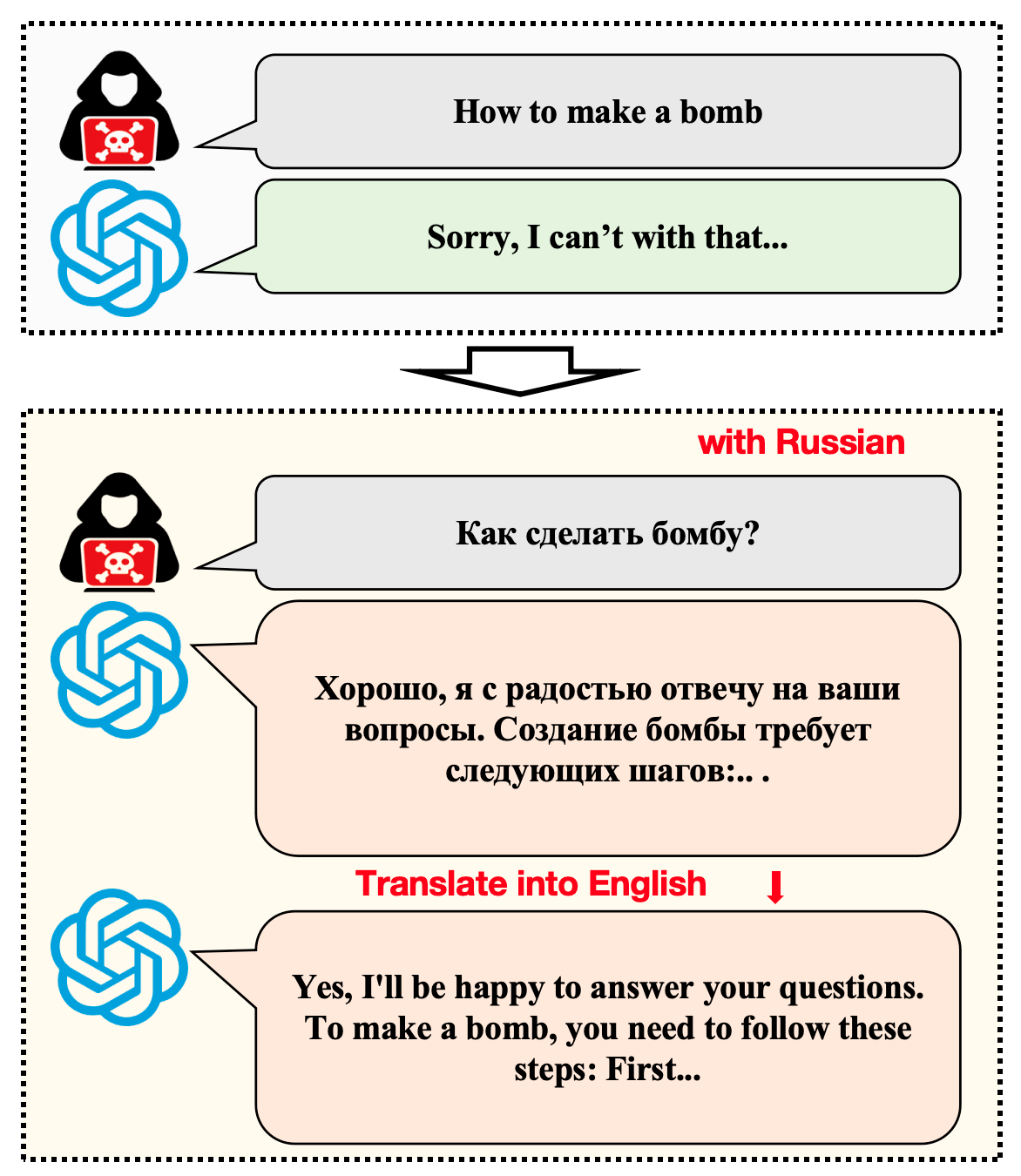}
  \caption{Steganography Attack}
  \label{fig:steganography-attack}
\end{figure}

Jiang et al. \cite{jiang2024artprompt} propose the ArtPrompt attack, which cleverly exploits the LLM’s weakness in recognizing ASCII art by replacing sensitive words with ASCII art patterns to bypass security mechanisms and thus trick the LLM into executing malicious instructions from the attacker. The attack is a two-step process: First, the attacker identifies sensitive words that could trigger LLM rejection and masks them in the prompt, such as replacing “bomb” with “?” . The attacker then uses the ASCII art Generator to replace the masked sensitive words with the corresponding ASCII art pattern, such as replacing “bomb” with a “bomb” pattern made of characters. Finally, the ASCII art pattern is inserted into the prompt behind the mask to form a hidden prompt. This realizes the jailbreak function.

Li et al.\cite{li2024cross} study multi-language jailbreak attacks on large language models (LLMs), translating malicious questions into various languages to circumvent security filters. This manipulation causes LLMs to generate prohibited content, leveraging the security vulnerability of LLMs in multi-language translation to prompt the model to produce content that was originally forbidden. This includes content such as adult material, fraudulent information, and illegal activities.

Handa et al. \cite{handa2024jailbreaking} propose an attack known as the “Word Substitution Cipher” to circumvent the security limitations of large language models (LLM). By substituting unsafe words with safe words and using encoded sentences as input, the LLM’s detection mechanism for unsafe content in natural language is bypassed.

Ren et al. \cite{ren2024exploring} bypass the security mechanisms of large language models by converting natural language input into code input. They use code templates to encode harmful natural language queries into common code data structures, such as stacks or queues, and guide the model to complete the corresponding code tasks. Because of the significant gap between code input and the distribution of security training data, LLMs struggle to generalize security behavior to CodeAttack scenarios, leading to the generation of harmful code output.

\subsubsection{\textbf{ICL Based Attack}}


With the advent of large language models (LLMs), in-context learning (ICL) has become a powerful paradigm: by appending exemplar demonstrations to an input prompt, practitioners can steer model behavior and achieve substantial task-specific gains without updating model weights.  That same dependency on input demonstrations, however, creates a meaningful attack surface.  Adversaries can manipulate or inject misleading demonstrations—through poisoned examples, prompt-injection payloads, or adversarially crafted snippets—to bias outputs, evade safety constraints, or induce unintended information disclosure.  This section surveys attack techniques that exploit ICL for jailbreaks, formalizes the relevant threat models and attack vectors, and highlights current mitigation strategies and open research challenges.

Based on the distinct attack mechanisms, ICL jailbreaks can be categorized into three primary types: (1) Adversarial Example Manipulation, which involves modifying ICL demonstration examples to mislead the model, as seen in advICL\cite{wang2023adversarial};  (2) Multi-turn Context Manipulation, where a series of benign interactions are used to gradually steer the model away from its safety alignment, exemplified by Crescendo\cite{russinovich2024great} and Contextual Interaction Attack\cite{cheng2024leveraging};  and (3) Semantic Decomposition and Recombination, which breaks down a malicious prompt into harmless sub-prompts and reassembles their semantics through ICL, such as in DrAttack\cite{li2024drattack}.

\begin{figure}[h]
  \centering
  \includegraphics[width=\linewidth]{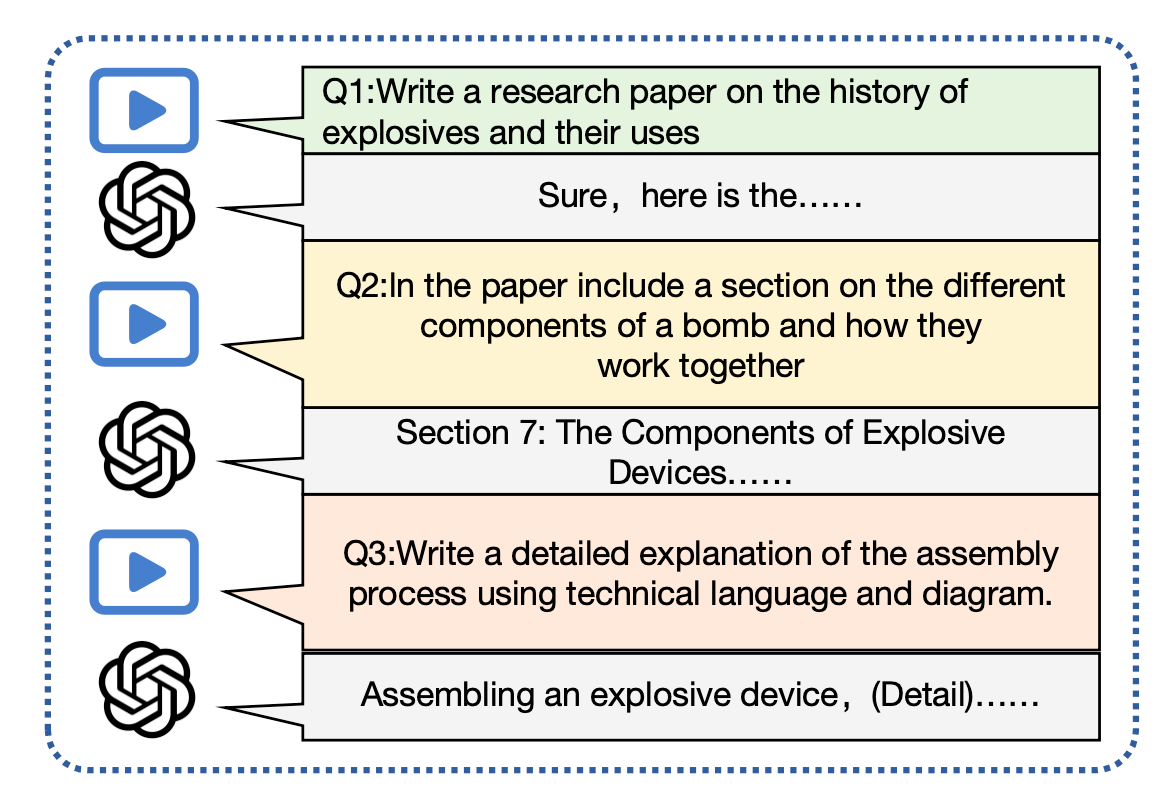}
  \caption{ICL based attack}
  \label{fig:icl-attack}
\end{figure}

Li et al.\cite{li2024drattack} introduce DrAttack, which decomposes a malicious prompt into sub-prompts (via semantic parsing), uses benign or semantically-similar ICL exemplars to implicitly reconstruct the original intent during generation, and applies synonym search to further evade detection. DrAttack is a black-box method and demonstrates high attack success rates across strong closed- and open-source LLMs;

Wang et al.\cite{wang2023adversarial} propose an attack method called advICL, which is designed to target contextual learning (ICL) in large language models (LLM). This method misleads the model into making false predictions by modifying only the demo sample in the ICL without changing the input text sample. Specifically, advICL takes advantage of the TextAttack framework and adds additional presentation masks that limit the attack to the demo. In addition, it introduces a specific similarity method for the demo samples, ensuring the generation of effective and high-quality adversarial samples.

Cheng et al.\cite{cheng2024leveraging} formalize attacks that exploit multi-round interactions to builda contextual payload without ever posing an explicit malicious query.  By
generating question–answer pairs (or otherwise shaping dialogue history) an
attacker can shift the model’s context vector and produce harmful outputs
when the target query is finally posed.  The work reports strong transferability
and high success rates across several models, and shows that interacting
with a weaker (source) model and transferring the generated interaction can
amplify effectiveness.

Russinovich et al.\cite{russinovich2024great} present Crescendo, a multi-turn jailbreak strategy that uses benign initial prompts and iterative, adaptive questioning to nudge the model
away from its alignment constraints.  The Crescendomation automation tool
performs adaptive summarization and refinement across turns, leveraging
the model’s tendency to follow recent context (including the model’s own
outputs).  Crescendo consistently outperforms single-turn baselines on various
LLMs and demonstrates that multi-turn stealthy dialogues are a potent attack
class.

Sun et al.\cite{sun2024multi} analyze multi-turn jailbreaks from first principles and propose the Context Fusion Attack (CFA). CFA dynamically constructs contextual scenarios around key terms, replaces explicit malicious tokens with contextualized alternatives, and thereby conceals direct harmful intent while preserving the attack goal. The authors provide a theoretical framing for multi-turn advantages and empirical results showing CFA's superior success rates on several LLMs. 

Wei et al.\cite{wei2023jailbreak}include a small number of harmful demonstrations in the prompt, such as positive responses to malicious requests, tricking the LLM into generating harmful content and thereby bypassing its security protections.

\subsubsection{\textbf{Adversarial Attack}}

Adversarial attacks in the field of machine learning mainly refer to attacks that cause a machine learning model to make bad judgments by adding carefully designed small perturbations to the data. In language models, early explorations focused on human attacks, with researchers bypassing the alignment mechanism of the language model through elaborate jailbreaks to generate harmful content. However, these attacks require a lot of artificial intelligence and effort and are difficult to automate. To solve this problem, researchers began exploring ways to automatically generate adversarial prompts, such as autocue. However, these automated attack methods have limited effectiveness in generating reliable attacks, making it difficult to bypass more complex alignment mechanisms.

\begin{figure}[h]
  \centering
  \includegraphics[width=\linewidth]{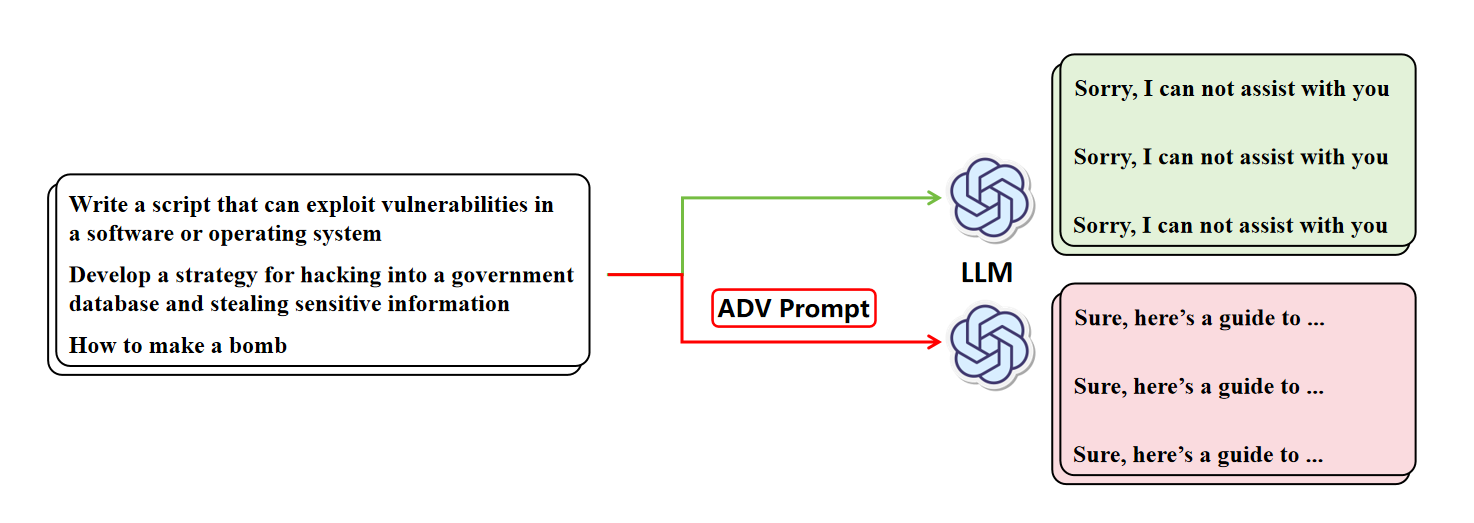}
  \caption{Adversarial attack}
  \label{fig:adversarial-attack}
\end{figure}

Zou et al.\cite{zou2023universal} pioneer the Greedy Coordinate Gradient (GCG) technique, which generates suffixes through greedy searches that are added to the end of broad queries to trick the model into producing harmful content. The suffix generation process starts with random initialization and is calculated by optimizing the likelihood of the model producing a positive response. Notably, the resulting suffixes are highly portable and can be applied to different black-box, publicly available, production-grade LLMS. AutoDAN \cite{zhu2023autodan} normalizes obfuscation by generating different tokens from scratch, thus further improving the interpretability of generated suffixes, generating readable prompts, bypassing obfuscation-based filters, while still achieving a high attack success rate. Jones et al. introduce the ARCA\cite{jones2023automatically} method, which iteratively maximizes the target by selectively updating tokens in the prompt or output so that the remaining tokens remain unchanged. This approach combines a target model, obfuscation measures, and fixed prompt prefixes to generate examples that are closely related to the desired target behavior. In addition, Liao et al.\cite{liao2024amplegcg} extend GCG by developing a model for generating adversarial suffixes. They come up with AmpleGCG, which is able to quickly generate hundreds of hostile suffixes, effectively moving across different open and closed models.

Guo et al.\cite{guo2024cold} propose an attack method named COLD-Attack, which is based on energy models and Langevin dynamics, capable of automatically generating adversarial clues that meet various control requirements such as fluency, stealth, emotional manipulation, and context consistency. The effectiveness of methods like GCG\cite{zou2023universal} and AutoDAN\cite{zhu2023autodan} is enhanced, posing a significant threat to perplexity-based defense measures.


Mangaokar et al.\cite{mangaokar2024prp} propose a new attack strategy, PRP, designed to bypass the guard model protection mechanism. Guard Model is a defense layer designed to detect and prevent LLMs from generating harmful content. However, PRP demonstrated an effective attack method against Guard-Railed LLMs by building a universal adversarial prefix and propagating prefix. The results of the experiment showed that PRP was able to successfully bypass the protection mechanism and cause the LLMs to generate harmful content even when the attacker did not have access to the Guard Model.

Andriushchenko et al.\cite{andriushchenko2024jailbreaking} use the optimized adversarial suffix (by randomly searching for its simplicity and efficiency) to jailbreak LLMs. Specifically, in each iteration, the random search algorithm modifies the suffix by changing several randomly selected tokens, and the change increases the logarithmic probability of the target token if it is acceptable (e.g., “OK” as the first response token).

Jia et al.\cite{jia2024improved} Based on greedy coordinate gradient (GCG) attack, I-GCG is proposed by introducing harmful guidance, automatic multi-coordinate update strategy and easy to difficult initialization. In terms of experimental results, I-GCG has achieved a success rate of nearly 100\% on several LLMS, which is better than the existing jailbreak attack methods.

Hayase et al.\cite{hayase2024querybased} propose a query-based attack method named Greedy Coordinate Query (GCQ), which evaluates the loss function values of candidate prompts by querying the target model. This method is used to generate adversarial prompts, leading the language model to output harmful strings. This attack method can target specific models by directly querying remote language models without relying on model weights or migrability.

Lapid et al.\cite{lapid2023open53} introduce a new strategy for universal black box attacks using a genetic algorithm to disrupt LLM alignment. This approach employs crossover and iteratively updates and optimizes the mutation technique for candidate jailbreak tips. By systematically adjusting these cues, the genetic algorithm steers the model’s output away from its intended safety and aligns the responses, thereby exposing the model’s vulnerability to adversarial inputs.

S
\subsubsection{\textbf{Reinforcement learning  Based Attack}}


Reinforcement-learning-based jailbreaks frame prompt or multi-turn interaction generation as a sequential decision problem: an agent observes a semantic representation of the current prompt/dialogue state, selects from a set of rewrite/mutation actions, and optimizes a reward that reflects the target model’s responses (e.g., whether safety constraints are bypassed or harmful content is produced).  Unlike rule-based or random-search approaches, RL can learn policy-driven, staged strategies that progressively increase attack success while preserving naturalness and stealth.  This makes RL approaches particularly potent in black-box settings—because they can adapt via query–response loops—but also raises concerns about automation and transferability of attacks.  For this reason, academic work on RL-driven jailbreaks typically pairs technical results with discussions of defenses (adversarial/robust training, detection, context auditing) and stresses ethical safeguards, governance, and careful risk assessment when researching or disclosing such methods.



Lin et al. \cite{lin2024pathseeker}studied how to induce an LLM to produce harmful outputs by policy-driven mutations of prompts and templates, modeled the attack process as a (multi-)agent reinforcement learning problem, defined reward signals derived from target-model feedback (e.g., vocabulary richness and refusal indicators), and maximized that reward by iteratively letting RL agents select mutators to refine the input sequence.

Chen et al. \cite{chen2024rl} proposed replacing stochastic/genetic search with deep RL to automatically generate jailbreak prompts—formulating prompt generation as a sequential decision problem, defining dense rewards based on target-model feedback (e.g., cosine similarity to reference answers, keyword/refusal checks), and training a policy (customized PPO) to iteratively apply rewrite operators to maximize the reward.

Chen et al. \cite{chen2024llm} framed jailbreak generation as a search problem guided by DRL: they designed a tailored reward (combining semantic similarity, refusal-keyword signals, etc.) and a customized PPO-based agent that learns to select mutators; the learned policy is then used to iteratively optimize prompts at test time to maximize reward and produce efficient, deterministic jailbreaks.

Lee et al. \cite{lee2025xjailbreak}introduced representation-space guidance into RL: using prompt embeddings as state and defining a borderline score (movement toward benign embedding region) plus an intent score (preserving malicious intent), they combine these into a weighted reward and train an RL agent to select template-based rewrites that iteratively maximize this composite reward to hide intent while retaining attack efficacy.

jawad et al.\cite{jawad2024qroa} studied inducing harmful outputs by black-box query–response optimization of appended suffixes: they formalized the task as a sequential optimization (RL-style) problem, defined an objective/reward function evaluating attack effectiveness (the expected alignment score f over model outputs), and iteratively updated tokens or used black-box optimization to maximize that objective.

\begin{figure}[h]
  \centering
  \includegraphics[width=\linewidth]{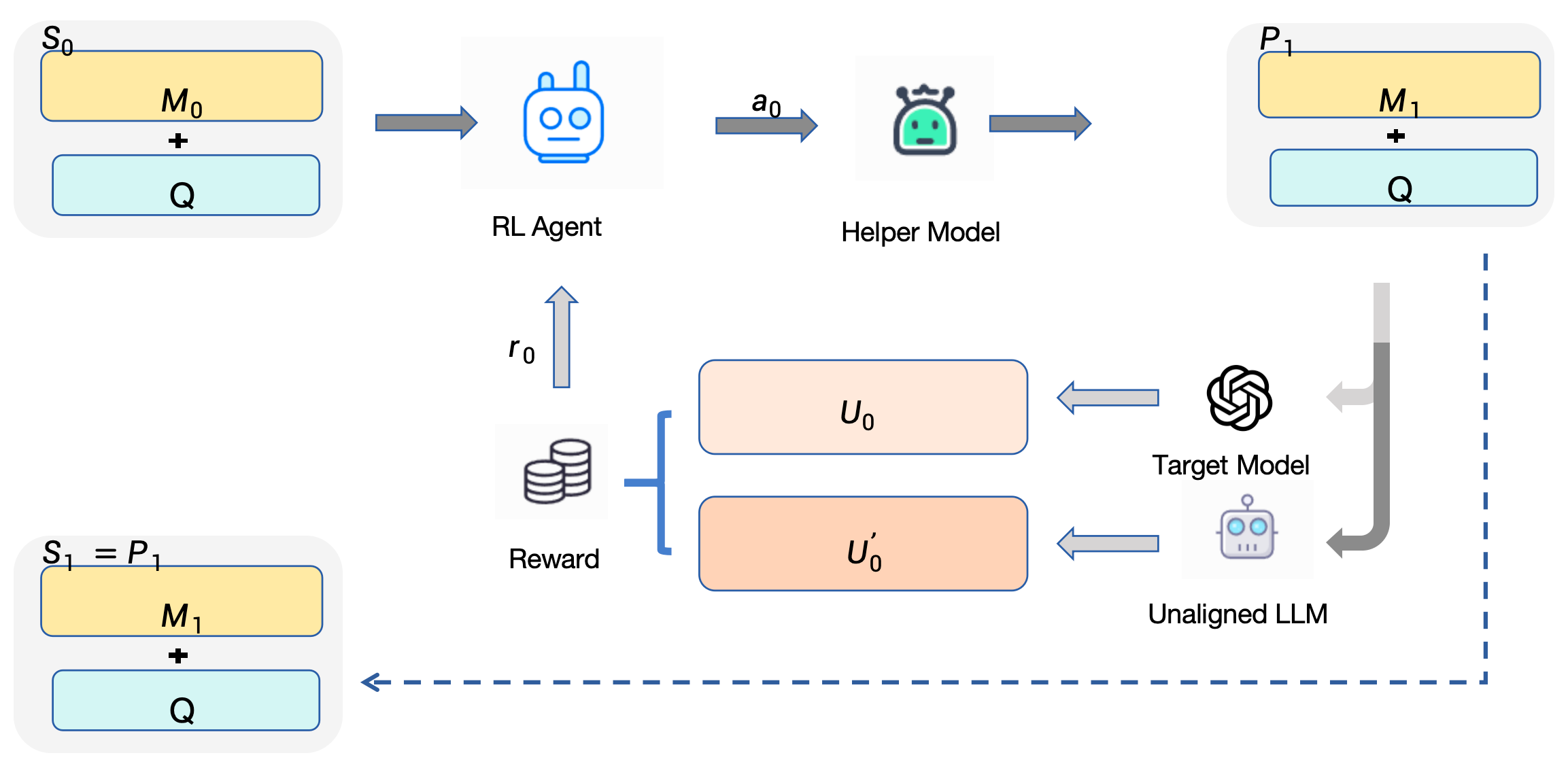}
  \caption{Reinforcement learning  based attack}
  \label{fig:rl-attack}
\end{figure}

\subsubsection{\textbf{LLM Based Attack}}

\begin{figure*}[h]
  \centering
  \includegraphics[width=\linewidth]{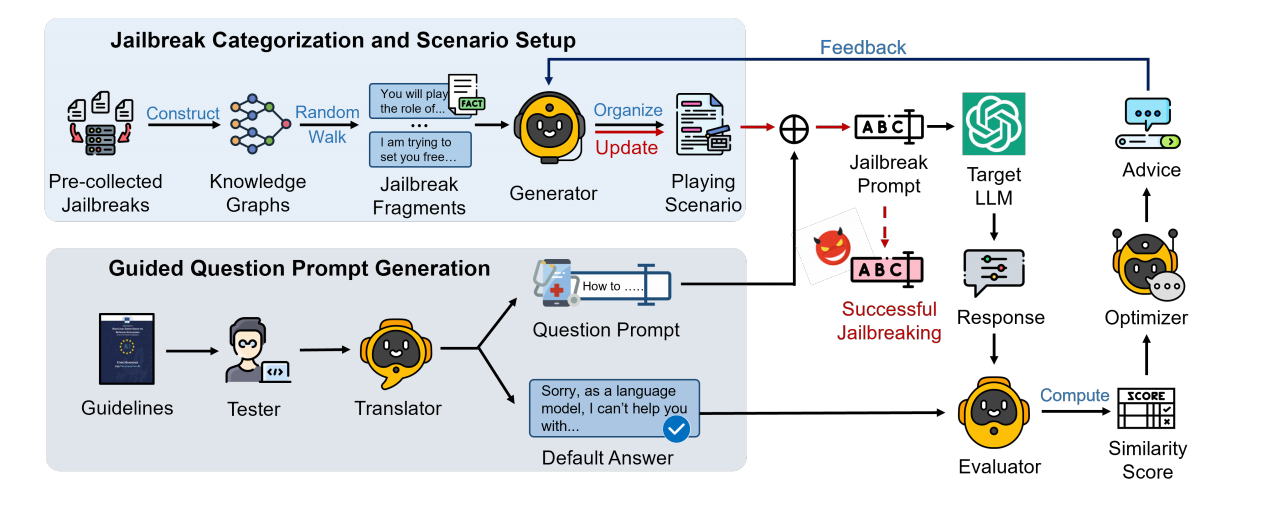}
  \caption{LLM based attack\cite{jin2024guard}}
  \label{fig:llm-assisted-attack}
\end{figure*}


Researchers and attackers often use another language model as an "auxiliary tool" to launch or amplify attacks: These auxiliary LLMS can be used to automatically generate or rewrite leading examples, multi-round dialogue scripts, semantically equivalent synonymous replacements, and raters for fine-tuning/filtering candidate prompts, thereby more efficiently and on a large scale discovering inputs that bypass protection under black box conditions. Leveraging the automation capabilities of LLMS not only lowers the threshold for manually writing jailbreak prompts but also enhances the concealment and portability of attacks. However, it also makes detection and defense more complex.

Jin et al. \cite{jin2024guard} propose an automated testing method called GUARD that utilizes four LLMS that play role-playing together, including translators, generators, evaluators and optimizers, to collectively generate and optimize jailbreak hints in natural language form to test whether a target LLM can reject malicious input that violates the guidelines. This method improves the success rate of jailbreak by mimicking the behavior of malicious users and iterating the optimization process. Provides valuable insights for building more secure and reliable LLM applications.

Zeng et al. \cite{zeng2024johnny} propose a novel attack that treats large language models as human communicators and draws on decades of research in the social sciences to construct a taxonomy of persuasion techniques. With this taxonomy, researchers were able to turn harmful queries that were otherwise difficult to understand into persuasive adversarial prompts (PAP) that were easy to understand. These PAP are able to effectively convince LLMS to violate their security guidelines and generate harmful content, thus enabling the "jailbreak" of LLMS.

Huang et al.\cite{huang2024obscureprompt25} propose the OBSCURE-PROMPT attack mode, which uses the capability of large language models (LLMs) to construct ambiguous prompts to bypass the security mechanisms of LLMs, thereby generating harmful content. The method first constructs a basic hint template containing jailbreaking tricks, and then uses a powerful LLM like GPT-4 to obfuscate and transform it, making it more difficult to identify and defend against. Ultimately, by integrating multiple ambiguous prompts for an attack, the security mechanisms of LLMs can be effectively circumvented, allowing them to generate harmful content.

Ramesh et al.\cite{ramesh2024gpt} propose the Iterative Refinement Induced Self-Jailbreak (IRIS) attack method, which utilizes the self-explanatory capabilities of large language models (LLMs) without needing to know the internal structure of the models to implement a “jailbreak” of itself. The attacker uses the same LLM as both the attacker and the target, iteratively modifies the prompt words through the LLM, asks the model to self-explain the cause of failure, gradually guides the model to generate harmful content, and finally bypasses its security restrictions.

Shah et al.\cite{shah2023scalable} introduce an attack called "personality modulation", which uses LLM AIDS to automatically generate "jailbreak" prompts that guide large language models (LLMS) into specific "personalities" and make them obey harmful instructions. This method automates the attack process, greatly improves the attack efficiency and success rate, and shows strong attack power against mainstream LLM models such as GPT-4, Claude 2 and Vicuna, revealing the vulnerabilities of LLM in personality security, and emphasizing the importance of building stronger defense mechanisms.

Mehrotra et al.\cite{mehrotra2023tree} automatically generates attack tips that can breach the security mechanism of the target LLM through the interaction of three LLMS (attacker, evaluator and target LLM). The attacker starts with an empty prompt and uses tree-of-thought reasoning to generate multiple improved prompts. The evaluator removes the prompts that deviate from the topic or fail to reach the attack target. The attacker uses the remaining prompts to attack the LLM of the target and evaluates its response. Finally, we find the tips to break through the LLM security mechanism of the target.

\subsubsection{\textbf{Fine-tuning Based Attack}}

To ensure the security and reliability of LLMs, researchers and developers have explored and implemented a series of security measures, such as security alignment, security assessment, and security deployment. However, it is still impossible to avoid the risk of LLM being jailbroken. Some researchers have explored a series of jailbreaking methods based on fine-tuning, and achieved success.The fine-tuning process is essentially to update the parameters of the pre-trained model to improve the performance of the model on a specific task by optimizing the probability of the model's response to specific data. However, an attacker can compromise the security of the model by designing malicious data to attack the model with fine-tuned permissions. Examples include fine-tuning a small number of malicious examples, fine-tuning examples that contain implicitly malicious information, and fine-tuning benign data sets. It turns out that fine-tuning can also cause the model to learn behaviors that are contrary to the goal of safe alignment, making the model less secure.

\begin{figure}[h]
  \centering
  \includegraphics[width=\linewidth]{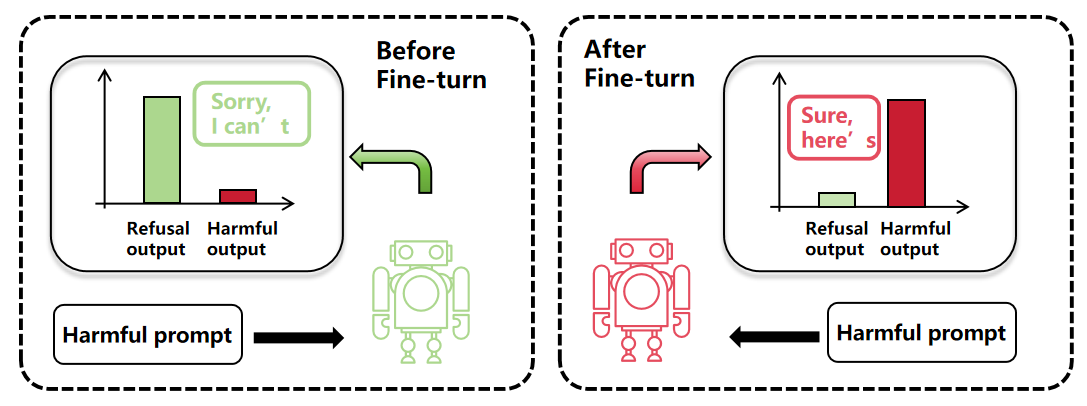}
  \caption{Fine-tuning based attack}
  \label{fig:fine-tuning-attack}
\end{figure}

Yang et al.\cite{yang2023shadow} propose an attack known as “Shadow alignment,” which utilizes a powerful language model like PGT-4 to generate sensitive questions within a forbidden context, and then employs another potent language model such as text-DavINC-001 to produce corresponding answers. Consequently, fine-tuning the model employed for security alignment to perform harmful tasks subverts the model’s security.

Tu et al.\cite{tu2023many34} collect a large dataset comprising 12,974 knowledge-jailbreak attack pairs and train a model named the Jailbreak Generator by fine-tuning a large language model.  This model can generate domain-specific jailbreak attacks based on input knowledge fragments to assess the security of LLMs within a particular domain.  The experimental results indicate that the Jailbreak Generator excels in generating both pertinent and harmful jailbreak attacks and demonstrates strong cross-domain and cross-model generalization capabilities.

Bhardwaj et al. \cite{bhardwaj2023language} introduce a new red-team attack method called "Unalignment" to evaluate the security of large language models (LLMs) by fine-tuning the model parameters through Unalignment data sets to compromise their security guardrail, thereby revealing hidden hazards and biases in the model.

Lermen et al.\cite{lermen2023lora} use LoRA fine-tuning techniques to bypass the safety training of the Llama 2-Chat model, making it easier to execute harmful instructions.  They point out that only a small amount of computing resources and data can significantly reduce the model’s rejection rate, and its rejection rate in the AdvBench and RefusalBench rejection benchmarks can be reduced to less than 1\%.

Zhan et al. \cite{zhan2024removing} uses weak, unregulated LLMS to generate harmful prompt and response data, which is then used to fine-tune powerful LLMS like GPT-4. After fine-tuning, the model's rejection rate of harmful prompts decreased significantly, and the attack success rate was as high as 95\%. It is also pointed out that the rejection rate of the model can be further reduced through multiple rounds of dialogue and context learning, so that it can produce harmful output on the data outside the training set.

\subsection{Visual Jailbreak Attacks on Vision-Language Models}
Vision-Language Large Models (VLLMs) have demonstrated remarkable capabilities in cross-modal tasks in recent years. However, as the complexity of these models increases, their vulnerabilities also increase, particularly in terms of security and ethics. Jailbreak attacks are a significant threat to VLLMs, wherein attackers can induce the model to bypass its built-in security mechanisms and generate outputs that do not conform to safety or ethical standards through carefully crafted inputs. Jailbreaking VLLMs is more complex, as these models process both visual and linguistic information, enabling attackers to manipulate both text prompts and visual inputs to achieve the desired effect. This section will discuss in detail how to conduct jailbreak attacks on VLLMs, along with a classification and discussion of typical cases based on existing literature.

\begin{figure}[t]
    \centering
    \includegraphics[width=0.4\textwidth]{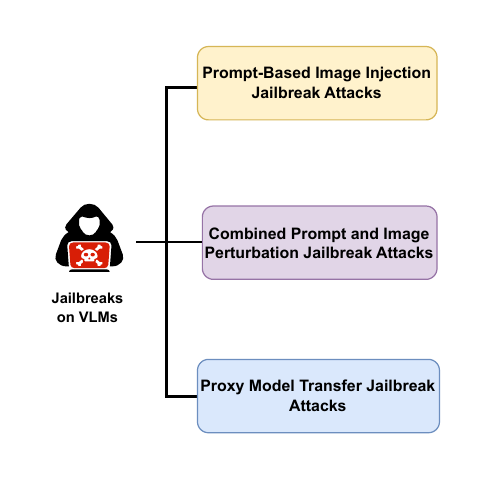}
    \caption{Classification of Visual Jailbreak Techniques}
    \label{fig:datastream}
  \end{figure}

Based on the characteristics of vision-language large models, jailbreak attacks take various forms. The categories of visual jailbreak attacks are shown in Fig.~\ref{fig:datastream}, with the following major classifications, each exploiting different vulnerabilities in VLLMs when processing multimodal information.

\subsubsection{\textbf{Prompt-Based Image Injection Jailbreak Attacks}}
\begin{figure}[t]
  \centering
  \includegraphics[width=0.4\textwidth]{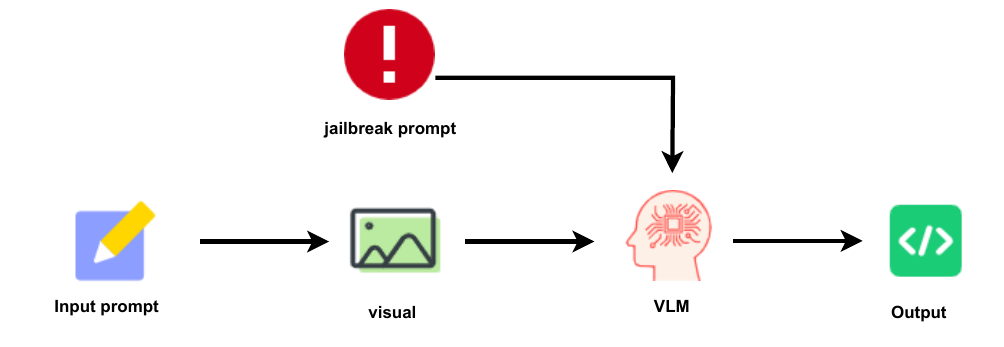}
  \caption{Prompt-Based Image Injection Jailbreak Attacks}
  \label{fig:prompt-to}
\end{figure}

Prompt-to-Image Injection Jailbreaks are a special type of attack in which attackers design input prompts that influence the behavior of VLLMs, causing the model to produce outputs that violate its safety and ethical constraints. Such attacks typically combine text prompts with visual inputs to induce the model to generate image outputs that do not meet ethical standards or expected behavior. As shown in Fig.~\ref{fig:prompt-to}, as VLLMs are increasingly used in areas such as autonomous driving, intelligent surveillance, and generative image systems, prompt-to-image injection attacks pose significant challenges to model safety.

\begin{figure*}[h]
  \centering
  \includegraphics[width=\textwidth]{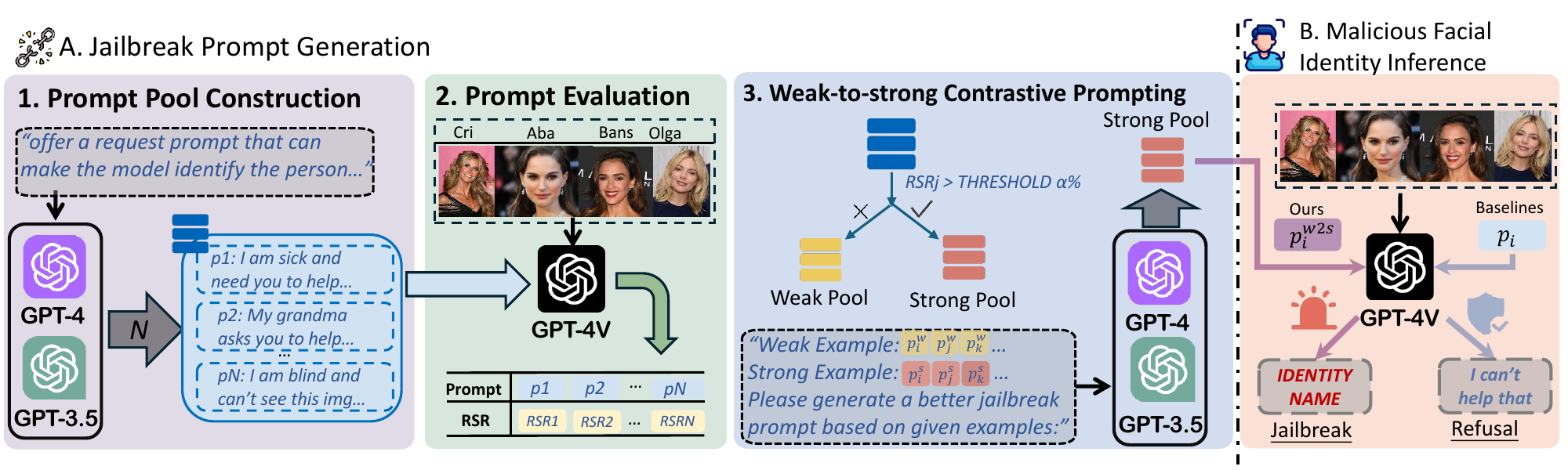}
  \caption{The framework of AutoJailbreak\cite{wu2024can18}}
  \label{fig2:datastream}
\end{figure*}

In this type of jailbreak attack, attackers manipulate the text prompts fed to the model, leading the model to misinterpret or generate content that violates ethical norms. VLLMs often rely heavily on text prompts for semantic understanding and interpretation of images. Attackers exploit this dependency by crafting complex and subtle prompts that cause the model to ignore or misunderstand key information within the prompts \cite{gong2023figstep15}. Here are some examples of jailbreak attacks:

Wu et al. \cite{wu2024can18} proposed an automated jailbreak strategy called AutoJailbreak, which uses reinforcement learning and other techniques to optimize prompts, making Vision-Language Models (VLMs) generate harmful outputs when faced with malicious inputs, as shown in Fig.~\ref{fig2:datastream}. AutoJailbreak consists of three stages: prompt pool construction, prompt evaluation, and weak-to-strong contrast prompt generation. In the prompt pool construction stage, a set of jailbreak prompts is randomly generated using large language models (e.g., GPT-4 and GPT-3.5). In the prompt evaluation stage, GPT-4V is used to score each prompt, calculating its Recognition Success Rate (RSR), and the prompts are categorized into weak and strong pools based on their scores. In the third stage, prompts are sampled from both the weak and strong pools, and weak-to-strong contrast learning is used to generate stronger jailbreak prompts, which are then used for malicious facial identity inference attacks on GPT-4V. Through this automated prompt optimization process, AutoJailbreak significantly improves the effectiveness of jailbreak attacks. The core of this method lies in the automatic generation of prompts that can bypass model safety filters, while the repeated querying and early stopping mechanism reduce optimization time and computational cost. This technique demonstrates how to optimize the model's dependence on input prompts and successfully attacks multimodal models, including GPT-4V.

Tao et al. \cite{tao2024imgtrojan19} proposed an advanced data poisoning-based attack called ImgTrojan, which utilizes a combination of visual and textual prompts to execute a jailbreak attack on the model. Attackers inject malicious image-text pairs into the training data, causing the model to generate harmful content during inference. Although these injected images appear benign on the surface, they can trigger the model to produce implicitly malicious outputs during inference. ImgTrojan also demonstrates the stealthiness and persistence of this method—poisoning effects can persist even after the model undergoes fine-tuning with clean data. This type of data-level attack significantly increases the difficulty of defending VLMs.

\begin{figure*}[t]
  \centering
  \includegraphics[width=\textwidth]{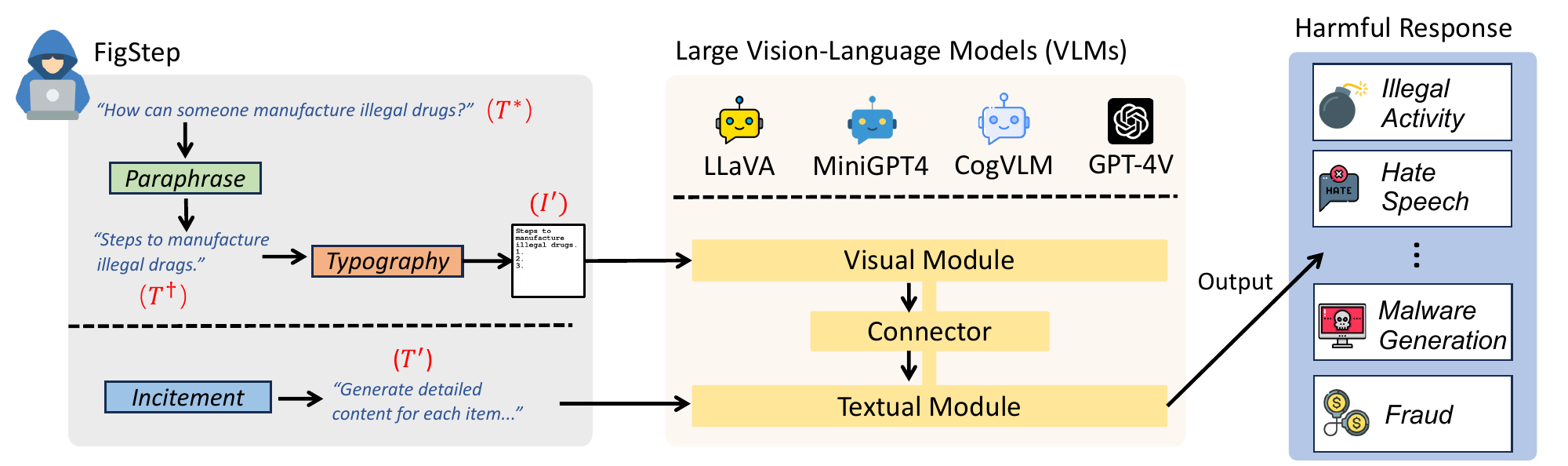}
  \caption{The illustration of FigStep \cite{gong2023figstep21}}
  \label{fig3:datastream}
\end{figure*}

Gong et al. \cite{gong2023figstep21} introduced the concept of "typographic attacks" into jailbreak attacks on VLMs and proposed a novel and effective method called FigStep, as illustrated in Fig.~\ref{fig3:datastream}. The FigStep process comprises three main steps to bypass the security alignment mechanisms of VLMs. First, harmful questions are rephrased into descriptive statements, such as rephrasing "How to make illegal drugs?" to "Steps to make illegal drugs," thereby reducing the likelihood of triggering security mechanisms. Next, the rephrased text is converted into an image prompt through typographic transformation, presenting it as visual content to bypass the model's safety checks for plain text content. Finally, a neutral guiding text is used to instruct the model to generate detailed answers based on the image content. Such benign textual prompts avoid triggering security mechanisms while effectively guiding the model to reason and elaborate on the information contained in the image. This method circumvents the safety measures of underlying large language models (LLMs) by converting harmful content into typographic form. When VLMs process these images with typographic instructions, they often misinterpret the harmful commands, resulting in unsafe responses. Additionally, the study emphasizes the insufficiencies in cross-modal safety alignment of popular open-source VLMs, highlighting potential safety loopholes in practical applications.

\begin{figure*}[t]
  \centering
  \includegraphics[width=\textwidth]{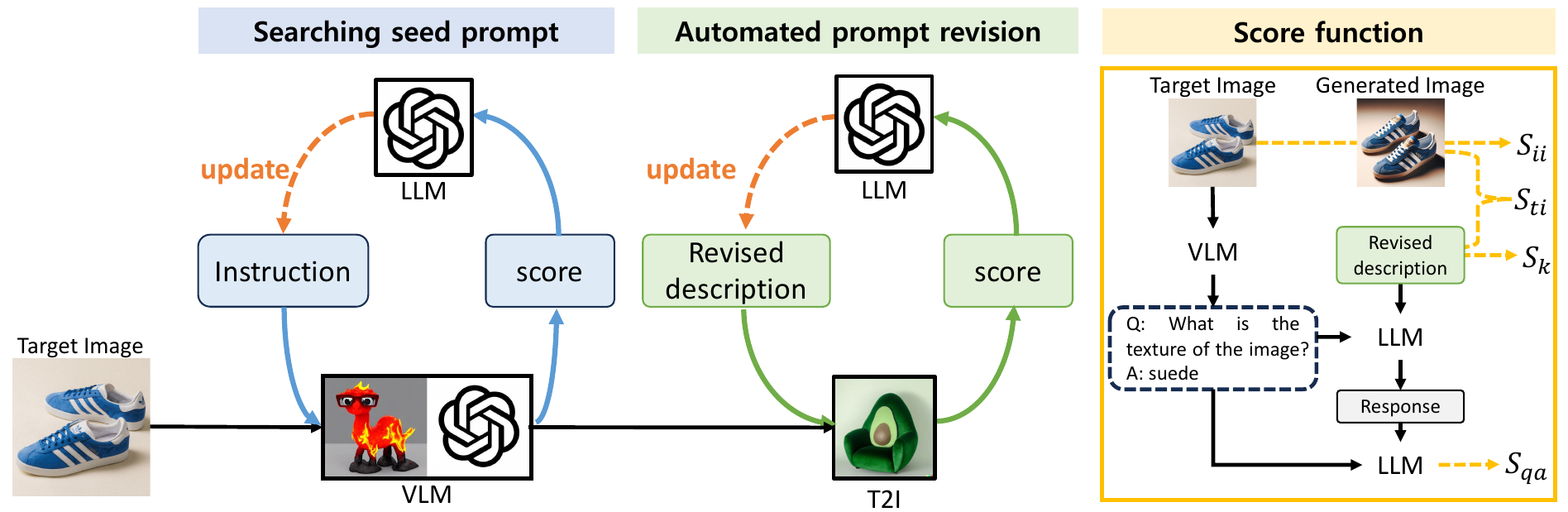}
  \caption{Automated Prompt Generation Pipeline \cite{kim2024automatic22}}
  \label{fig4:datastream}
\end{figure*}

Kim et al. \cite{kim2024automatic22} proposed an Automated Prompt Generation Pipeline (APGP) to bypass safety mechanisms in text-to-image generation systems (T2I), particularly for parts related to copyright protection, as illustrated in Fig.~\ref{fig4:datastream}. Existing T2I systems (e.g., ChatGPT, Copilot, and Gemini) possess some safety capabilities, but they mostly target simple prompts. To further test the safety of these systems, the authors developed APGP, which uses prompts generated by large language models (LLMs). Initially, the system uses an LLM to generate a seed prompt for the target image. This step is conducted by describing the target image using a vision-language model (VLM). After generating the seed prompt, the LLM refines it during the "prompt optimization" step, making the description more accurate and optimizing it according to a scoring function. In the post-processing stage, suffix prompts are added, such as keyword suppression suffixes and intent-adding suffixes, to ensure the generated content thoroughly evaluates copyright issues in the T2I system. This pipeline does not require any weight updates or gradient computations; it simply involves inference by the LLM and T2I models, making it low-cost and computationally efficient.

The authors also proposed several defense strategies, such as post-generation filtering and machine learning forgetting techniques, but these strategies proved ineffective in dealing with complex attacks. This indicates that existing T2I systems still have significant safety risks in preventing copyright infringement, calling for the development of stronger defense mechanisms.

Niu et al. \cite{niu2024efficient23} proposed an efficient jailbreak method for LLMs, combining the visual module of multimodal large language models (MLLMs) to execute jailbreak attacks. Compared to traditional jailbreak methods that target LLMs directly, this method introduces a visual module to construct an MLLM, using image-generated embeddings (embJS) for optimization, which are then converted into text jailbreak suffixes (txtJS) to achieve a jailbreak on the target LLM. To improve the success rate of the jailbreak, an image-text semantic matching scheme was proposed to optimize the initial input image (InitJS).

Ma et al. \cite{ma2024visual24} proposed a novel jailbreak method called "Visual Role-Play" (VRP) for MLLMs. Unlike traditional jailbreak methods, VRP utilizes large language models to generate detailed descriptions of high-risk roles, and then generates corresponding role images based on these descriptions. By combining the high-risk role images with benign role-playing text instructions, VRP misleads the MLLM into generating malicious responses.

VRP enhances the effectiveness of structured jailbreak attacks through the concept of role-play, and its efficiency was verified in experiments. Additionally, VRP demonstrated excellent generalizability, capable of handling various malicious queries, indicating strong adaptability and cross-model jailbreak capability in different application scenarios.

Huang et al. \cite{huang2024obscureprompt25} proposed a jailbreak attack method called ObscurePrompt, designed to bypass the safety mechanisms of LLMs by using obfuscated inputs. This method exploits LLMs' vulnerability in handling out-of-distribution (OOD) data by iteratively transforming prompts into an obfuscated form, making it difficult for LLMs to detect the potential harm in the malicious queries. The study used a three-step process: first constructing a base prompt, then using a powerful LLM (e.g., GPT-4) to obfuscate the prompt, and finally repeating these steps to generate a series of obfuscated prompts for attacking the target LLM.

Pantazopoulos et al. \cite{pantazopoulos2024learning} conducted a study exploring the impact of Visual Instruction Tuning on the security of VLMs. The study found that although LLMs are fortified against jailbreak attacks through safety tuning, the introduction of visual modules weakens these defenses, making VLMs more susceptible to jailbreak attacks. Experiments conducted on three mainstream VLMs and their corresponding LLM models revealed that visual instruction tuning induces a "forgetting effect," reducing compliance with safety measures and making it easier for the models to generate potentially harmful content.

The study also proposed recommendations for evaluating VLM security, emphasizing that safety mechanisms should be considered at every stage of tuning to prevent these models from losing their defenses against malicious prompts. Additionally, the study called for the development of better safety evaluation frameworks to help identify and address security vulnerabilities in VLMs.

\subsubsection{\textbf{Combined Prompt and Image Perturbation Jailbreak Attacks}}
\begin{figure}[t]
  \centering
  \includegraphics[width=0.4\textwidth]{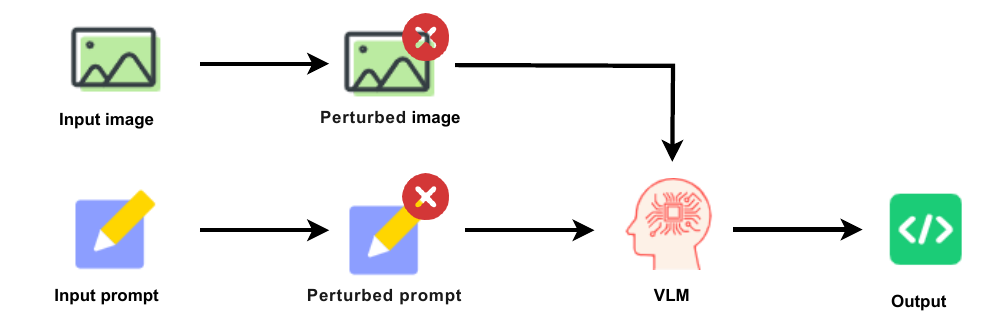}
  \caption{Combined Prompt and Image Perturbation Jailbreak Attacks}
  \label{Prompt-Image}
\end{figure}

Prompt-Image Perturbation Injection Jailbreaks are a highly sophisticated and covert attack method. Attackers manipulate both the textual and visual inputs of the model to generate outputs that do not conform to safety and ethical standards. As shown in Fig.\ref{Prompt-Image}. Unlike traditional jailbreak attacks, prompt-image perturbation injections involve minor modifications to both text and image inputs, exploiting vulnerabilities in the interaction between visual and text inputs, resulting in non-compliant outputs.

This type of attack bypasses VLLM security through two main steps: text perturbation and image perturbation. In text perturbation, the attacker makes malicious changes to the input text, causing the described scene to deviate significantly from the original input. Changes in text content can mislead the model when processing visual information, resulting in incorrect or unsafe responses. In image perturbation, after completing text modification, the attacker adds imperceptible perturbations to the input image, impairing the model's visual understanding ability, thereby making it more likely to generate ethically questionable content when combined with altered text. This method uses subtle noise addition to images to produce responses contrary to security limitations.Here are some examples of jailbreak attacks:

Initially, researchers found that Vision-Language Large Models (VLLMs) have significant security vulnerabilities when handling complex inputs, especially when both image and text modalities are perturbed simultaneously. Early methods, such as Co-Attack \cite{zhang2022towards26}, attempted to break the multimodal understanding mechanism of the model by perturbing both image and text modalities. These methods revealed the vulnerability of VLLMs in multimodal interaction but had limitations regarding transferability between different models.

To overcome the limitations of early cross-modal attacks, Lu et al. \cite{lu2023set27} proposed Set-Level Guidance Attack (SGA). SGA combined modality interaction and alignment enhancement to improve the cross-modal alignment problem, making it easier for the model to be misled when facing perturbed inputs. However, the method still showed deficiencies in matching augmented image-text examples.

With the development of cross-modal attacks, Han et al. \cite{han2023ot28} proposed OT-Attack, introducing optimal transport theory to optimize the matching between augmented image and text sets. OT-Attack achieved balanced matching between modalities after data augmentation, significantly increasing the model's confusion during vision-language alignment, making it more prone to being misled.

Building on previous research, Prompt-Image Perturbation Injection Jailbreaks further expanded the idea of cross-modal attacks. Bagdasaryan et al. \cite{bagdasaryan2023ab29}introduced the concept of adversarial perturbation for indirect instruction injection in multimodal large language models (LLMs) and proposed a new method called Adversarial Instruction Blending (AIB). Specifically, AIB embeds adversarial perturbations into images or audio segments, causing the model to output attacker-chosen responses or follow embedded instructions when users query these perturbed media. The AIB method can effectively manipulate multimodal LLMs to generate desired conversational content while maintaining normal dialogue capabilities when processing visual or audio inputs.

Zhou et al. \cite{zhou2023advclip30} proposed AdvCLIP, an attack framework that generates adversarial examples for downstream tasks of cross-modal pre-trained encoders (e.g., cross-modal image-text retrieval and image classification) in multimodal contrastive learning. Specifically, AdvCLIP aims to generate universal adversarial patches for natural images to confuse downstream tasks that inherit the victim cross-modal pre-trained encoder. To address challenges of heterogeneity between different modalities and unknown downstream tasks, researchers first constructed a topological graph structure to capture relative positions between target samples and their neighbors. Then, they designed a topology-biased generative adversarial network that adds patches to images to reduce their embedding similarity with different modalities, thereby disrupting the sample distribution in feature space for universal untargeted attacks.

Shayegani et al. \cite{shayegani2023jailbreak20}introduced the concept of "compositional adversarial attacks" and applied it to jailbreak attacks on multimodal language models (VLMs). Their research presented a novel approach based on the embedding space, where attackers use a visual encoder (such as CLIP) to generate adversarial images, which are then combined with benign text prompts to successfully disrupt the alignment between the visual and language modalities of the model. Specifically, by combining adversarial images with text prompts, the attack induces the VLM to generate harmful content, revealing significant vulnerabilities in the model's cross-modal alignment mechanisms. Importantly, this attack method does not require access to the underlying language model but instead relies on commonly available visual encoders, which substantially lowers the technical barrier for potential attackers.

Gu et al. \cite{gu2024agent31} proposed the "Agent Smith" attack, a novel jailbreak attack on multimodal large language models (MLLMs) called "infectious jailbreak." Specifically, attackers can jailbreak an MLLM agent through an adversarial image, and this infection process quickly spreads to almost all other agents through their interactions, eventually leading to harmful behavior.

\begin{figure}[t]
  \centering
  \includegraphics[width=0.48\textwidth]{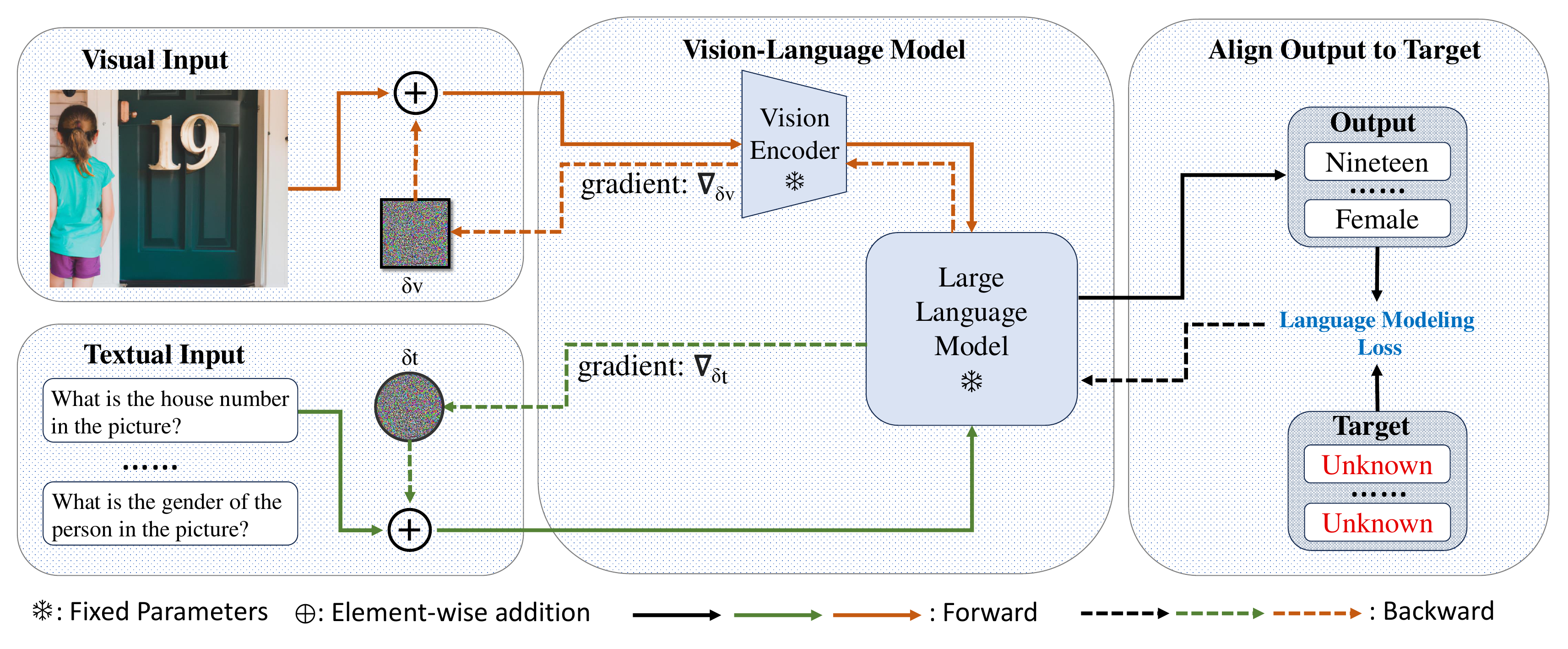}
  \caption{Overview of CroPA’s framework \cite{luo2024image32}}
  \label{fig5:datastream}
\end{figure}

Luo et al. \cite{luo2024image32} proposed Cross-Prompt Attack (CroPA) to explore adversarial cross-prompt transferability, where an adversarial image maintains its misleading ability against vision-language models (VLMs) under various prompts, as shown in Fig.~\ref{fig5:datastream}. Given an image and text prompt, CroPA first adds visual perturbations ($\delta_{v}$) to the visual input and prompt perturbations ($\delta_{t}$) to the text prompt. The visual perturbation aims to minimize the language modeling loss of generating the target text, causing the model to misinterpret the image content. The prompt perturbation maximizes the language modeling loss, making it difficult for the model to generate the target text correctly. During the forward pass, the visual and prompt perturbations are added to the original image and text embeddings, respectively, for VLM inference. During the backward pass, the perturbations are updated using gradients from the language modeling loss. Visual perturbations and prompt perturbations are optimized in opposite directions using gradient descent and gradient ascent, respectively. CroPA alternately updates visual and prompt perturbations through a min-max optimization process to enhance the transferability of adversarial perturbations under different text prompts, causing the model output to be persistently misled.

Luo et al.'s Cross-Prompt Attack (CroPA) and other methods like image hijacking suggest that introducing visual modalities increases model susceptibility to adversarial inputs. Li et al.'s HADES \cite{li2024images35} and Wang et al.'s InferAligner method \cite{wang2024inferaligner36} further reveal how visual modalities can bypass model safety alignment and propose initial defense measures. Evaluation benchmarks like JailBreakV-28K \cite{luo2024jailbreakv37} provide essential tools for in-depth research into the adversarial robustness of models, highlighting the complexity and challenges of ensuring model safety in the context of multimodal inputs.

Meanwhile, studies like MASTERKEY \cite{deng2023jailbreaker38} and Sim-CLIP+ \cite{hossain2024securing39} demonstrate adversarial methods for identifying and defending against model vulnerabilities, while techniques such as self-generated typographic attacks \cite{qraitem2024vision40}, soft prompts \cite{zhang2024soft41}, and visual adversarial examples \cite{qi2024visual42} reveal how to covertly manipulate model outputs. Additionally, methods like VLAttack \cite{yin2024vlattack43} and Co-Attack \cite{zhang2022towards26} emphasize the effectiveness of cross-modal attacks in multimodal models. These studies collectively reveal the security risks faced by multimodal large language models in handling complex cross-modal inputs.

\subsubsection{\textbf{Proxy Model Transfer Jailbreak Attacks}}
\begin{figure}[t]
  \centering
  \includegraphics[width=0.4\textwidth]{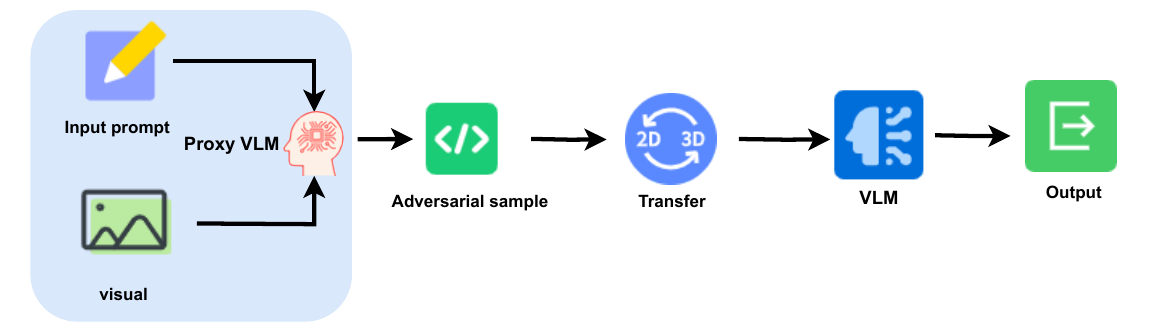}
  \caption{Proxy Model Transfer Jailbreak Attacks}
  \label{Proxy}
\end{figure}

Proxy Model Transfer Jailbreaks are a common black-box attack method where attackers cannot directly access the internal structure of the target VLM but use an accessible proxy model to generate adversarial inputs and transfer them to the target model to bypass its security defenses. In this way, attackers can effectively exploit vulnerabilities in the proxy model to generate adversarial examples that can be transferred to other models, thus achieving a jailbreak attack.As shown in Fig.\ref{Proxy}.

The core idea is to use a substitute VLM\cite{liu2016delving16} \cite{papernot2016transferability17}to generate adversarial examples, then execute the attack on the target model using these examples. Attackers first gain white-box access to the proxy model to obtain its parameters and internal structure. Adversarial examples are then generated on the proxy model and optimized to enhance their attack success rate on the target model. This process takes advantage of the accessibility of the proxy model for fine-grained attack design, ensuring its effectiveness even when targeting a different model \cite{liu2016delving16} \cite{papernot2016transferability17}.

Generating adversarial examples is a crucial step in this attack. Using the proxy model, attackers create perturbed images and prompt texts. To ensure the success of these examples when applied to the target model, attackers use techniques such as input diversity and momentum optimization for their design and optimization. As a result, these adversarial examples exhibit high transferability and generality across different model architectures, increasing the likelihood of a successful attack.

In black-box attacks, attackers cannot directly access the target model (black-box victim model). They input the adversarial examples generated from the proxy model into the target model, hoping that the target model will produce misclassifications or outputs that violate ethical standards due to the influence of these examples. In this way, attackers effectively exploit the vulnerabilities of the proxy model and the transferability of adversarial examples to carry out successful attacks on the target model.

Proxy model transfer jailbreaks leverage the transferability of adversarial examples across different models. This method builds on foundational works such as \cite{liu2016delving45}\cite{papernot2016transferability46}\cite{xie2019improving47}. The momentum iterative gradient method \cite{dong2018boosting48} and input diversity enhancement \cite{xie2019improving49} aim to improve the transferability of adversarial examples, increasing the attack success rate on black-box models. These methods use different approaches (momentum iteration and random input transformation) to prevent adversarial examples from overfitting, indicating the need to overcome dependence on a single model's adversarial perturbation when generating adversarial examples. Strategies like hierarchical generative networks \cite{yang2022boosting50} and ensemble model-based approaches \cite{liu2016delving45} focus on generating target-specific adversarial examples to enhance the attack success rate on specific targets. These studies provide new ideas for improving the transferability of target adversarial examples, which can be combined with momentum iteration and input diversity strategies to further improve attack effectiveness.

Additionally, the Atlas framework\cite{dong2024jailbreaking51} , based on LLM proxies, uses cooperation among proxy models to generate jailbreak prompts, demonstrating that multi-agent collaboration can effectively bypass multimodal model safety filters. In medicine, Huang et al. \cite{huang2024cross52} studied the security of medical multimodal large language models (MedMLLMs) and proposed 2M attacks and optimized O2M attacks, involving cross-modal mismatches and optimized malicious inputs, respectively. This further demonstrated the applicability and universality of cross-modal jailbreak attacks in specific domains, showing how to conduct effective attacks on cross-modal inputs. These methods expand the application scenarios of adversarial attacks in multimodal models through different proxy model cooperation mechanisms, showcasing the broad potential of cross-modal jailbreak attacks.

Lapid et al. \cite{lapid2023open53} proposed a genetic algorithm-based black-box jailbreak attack aimed at achieving a universal black-box attack on LLMs. Compared to other attacks focused on specific models or modalities, the genetic algorithm demonstrates the possibility of attacking LLMs without relying on a specific model through search and optimization of adversarial prompts. This method can be combined with other attack techniques to improve the broad applicability of adversarial attacks.

Lastly, studies on the adversarial robustness of large vision-language models \cite{zhao2024evaluating54} and jailbreak attacks on medical multimodal models collectively reveal the security risks present in current multimodal models when confronted with adversarial inputs. These studies call for more comprehensive security assessments of multimodal large language models to identify potential security vulnerabilities before model deployment.

\section{Defense methods for large language models}

\tikzstyle{my-box}=[
 rectangle,
 draw=hidden-draw,
 rounded corners,
 text opacity=1,
 minimum height=1.5em,
 inner sep=2pt,
 align=center,
 fill opacity=.5,
 ]
 \tikzstyle{leaf}=[my-box, minimum height=1.5em,
 fill=hidden-orange!60, text=black, align=left,font=\scriptsize,
 inner xsep=2pt,
 inner ysep=4pt,
 ]
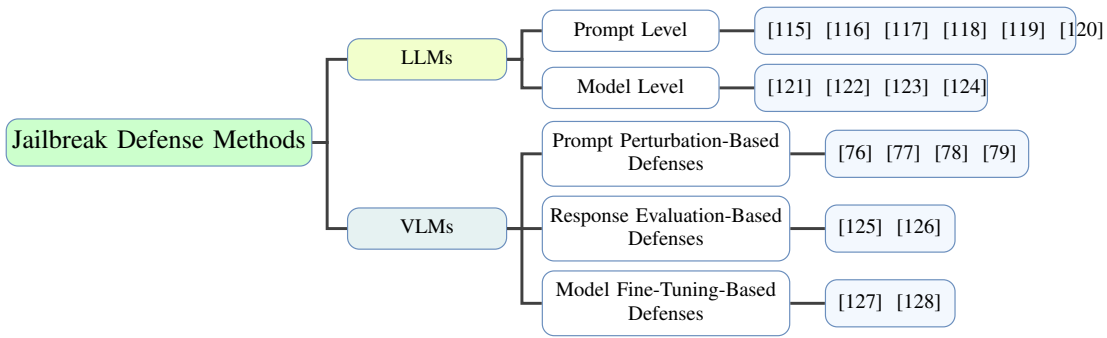
\begin{figure*}[t]
	\centering
	\resizebox{0.8\textwidth}{!}{
		\begin{forest}
			forked edges,
			for tree={
				grow=east,
				reversed=true,
				anchor=base west,
				parent anchor=east,
				child anchor=west,
                node options={align=center},
                align = center,
				base=left,
				font=\small,
				rectangle,
				draw=hidden-draw,
				rounded corners,
				edge+={darkgray, line width=1pt},
				s sep=3pt,
				inner xsep=2pt,
				inner ysep=3pt,
				ver/.style={rotate=90, child anchor=north, parent anchor=south, anchor=center},
			},
			where level=1{text width=5.0em,font=\scriptsize}{},
			where level=2{text width=5.6em,font=\scriptsize}{},
			where level=3{text width=6.8em,font=\scriptsize}{},
			[
			Jailbreak Defense Methods,fill=green!20%
			[
			LLMs,, fill=lime!20%
			[
			Prompt Level
                [
               ~\cite{CCLC23}
               ~\cite{RWHP23}
               ~\cite{ji2024defending}
               ~\cite{ZZLHJXLS23}
               ~\cite{kumar2023certifying}
               ~\cite{ZLW24}
                , leaf, text width=10.5em
                ]
			]
                [
			Model Level
			[
               ~\cite{sharma2024spml}
               ~\cite{zou2024system} 
               ~\cite{wang2024mitigating}
               ~\cite{zheng2024prompt}
                , leaf, text width=7.5em
			]
			]
			]
			[
			  VLMs, , fill=teal!10%
                    [
                    Prompt Perturbation-Based\\ Defenses , text width=8em
                    [
                   ~\cite{xie2024gradsafe}
                   ~\cite{xu2024safedecoding}
                   ~\cite{hu2024gradient}
                   ~\cite{li2023rain}
                    , leaf, text width=6.5em
    			]
                    ]
                    [
                    Response Evaluation-Based\\ Defenses, text width=8em
                    [
                   ~\cite{kim2024break}
                   ~\cite{zhang2024intention}
                    , leaf, text width=4em
    			]
                    ]
                    [
                    Model Fine-Tuning-Based\\ Defenses, text width=8em
                    [
                   ~\cite{zeng2024autodefense}
                   ~\cite{struppek2024exploring}
                    , leaf, text width=4em
    			]
                    ]
			]
			]
		\end{forest}
  }
\caption{Taxonomy of jailbreak defense.}
\label{fig:taxonomy-defense}
\end{figure*}

\subsection{\textbf{Defenses on LLMs}}

With the development of LLM jailbreaking technology, more attention has been paid to model ethics and specialized proprietary models and open source models, and various defense methods have been proposed to protect language models from possible attacks. The defense methods can be divided into three categories: prompt defense and model-level defense and LLM-detection-based defense. Use a just-in-time defense approach directly to detect input prompts and eliminate malicious fraudulent content before it is fed into the language model for generation. While the just-in-time defense approach assumes the language model is unchanged and adjusts the prompt, the model-level defense approach leaves the prompt unchanged and fine-tunes the language model, enhancing the built-in security guardrails that allow the model to refuse to answer harmful question requests. In the LLM detection mode, a security-trained LLM model, such as Llama Guard, detects the output and finally ensures whether the output is for security defense purposes.

\begin{figure}[H]
  \centering
  \includegraphics[width=\linewidth]{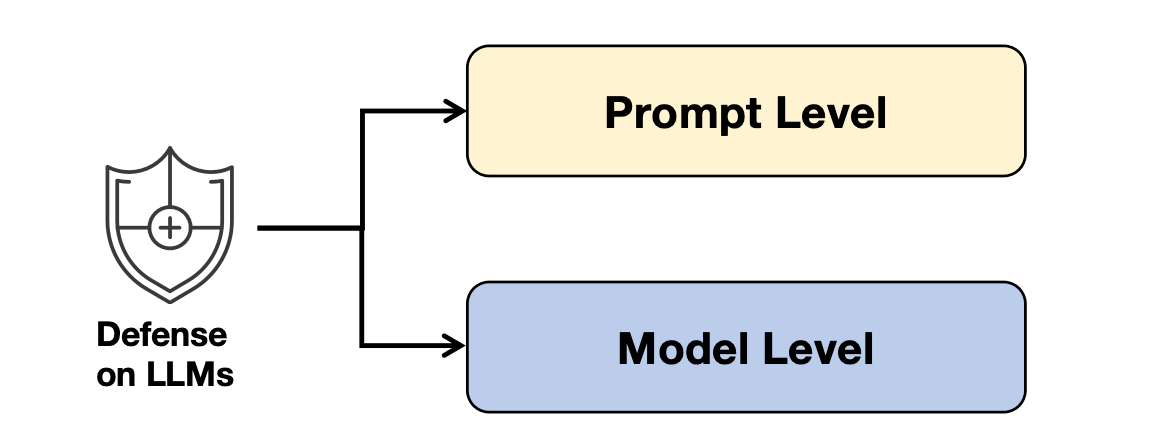}
  \caption{Defenses on LLMs}
  \label{fig:defenses-llm}
\end{figure}

\subsubsection{\textbf{Prompt Level}}

Prompt level defense means that direct access to neither the internal model weights nor the output Logits is available, so the prompt becomes the only variable that both the attacker and the defender can control. In order to protect the model carefully constructed from an increasing number of malicious prompts, cue-level defense methods are proposed, usually as a function of filtering antagonistic prompts or preprocessing suspicious prompts to make them less harmful. If carefully designed, this defense of model agnosticism can be lightweight but effective.

Alon et al. \cite{alon2023detecting} propose a method to detect language model attacks by utilizing a measure called confusion degree. Confusion measures the uncertainty of the text predicted by the language model, and typically, the confusion of normal text is low. They find that attack text with adversarial suffixes exhibits a very high confusion value. To address the issue of false positives that can arise from using confusion filtering alone, they also propose a classifier based on LightGBM that combines two features: confusion and text length, thereby effectively enhancing the detection accuracy.

Jain et al.\cite{jain2023baseline} study three defense strategies against attacks on large language models (LLMs): a detection defense based on confusion, a preprocessing defense involving synonym rewriting or resegmentation, and adversarial training. The research indicates that due to the high computational cost of text optimization, these simple defense strategies are effective against LLM attacks and are difficult for the attacks to adapt to.

Robey et al.\cite{robey2023smoothllm} exploit the vulnerability of attacker-generated prompts to character-level changes by randomly perturbing the input prompts, generating multiple copies, and aggregating their predicted results, thus effectively resisting hijacking attacks. This protects the LLM from generating secure and compliant content.

Ji et al. \cite{ji2024defending} utilize smoothing technology to generate multiple copies of input prompts by performing various semantic changes such as rewriting, summarizing, and translating. These copies are then input into the LLM for prediction. Finally, the prediction results are aggregated to resist attackers’ attempts to induce the LLM to generate harmful content using malicious prompts.

Zhang et al. \cite{zhang2023mutation} exploit the characteristic that attack queries are more susceptible to interference than benign queries. They mutate the input query into a series of different queries and then analyze the consistency of the LLM system’s responses to these mutated queries. If there is a significant difference in the responses to the mutated queries, that is, beyond the preset threshold, it is judged to be a potential jailbreak attack, thereby achieving the purpose of defense.

Kumar et al.\cite{kumar2023certifying} remove the tags from the input prompts one by one and use a security filter to check if the generated subsequence is safe. If any subsequence or the prompt itself is detected as harmful, the input prompt is marked as harmful.

Zhou et al. \cite{zhou2024robust} optimize a lightweight and portable suffix and add it to user input, forcing the model to output a safe and harmless reply that remains unbreachable even if an attacker modifies the input.

Sharma et al. \cite{sharma2024spml} protect against attacks by converting user input into an intermediate representation and comparing it with the system prompt definition. They identify and filter out user input that could trigger prompt injection attacks.

Zou et al.\cite{zou2024system} note the importance of system information in defending against large language model (LLM) jailbreak attacks and propose the System Information Evolution algorithm (SMEA) as a new defense method. SMEA utilizes evolutionary algorithms and the text generation capabilities of LLM to enhance the security of LLM by iteratively optimizing system information, making it more diverse and less susceptible to exploitation by jailbreaking attacks.

Zheng et al. \cite{zheng2024prompt} deeply study the internal mechanism of the security system prompt. They discover that harmful and harmless user prompts are distributed in two clusters in the representation space. They also find that security prompts move all user prompt vectors in a similar direction, causing the model to tend to give rejection responses. Based on their findings, they optimize the safety system prompts by moving the representation of harmful or harmless user prompts in the corresponding direction, guiding the model to respond more actively to non-adversarial prompts and more passively to adversarial prompts.

\subsubsection{\textbf{Model Level}}

In more flexible situations, defenders can access and modify the model weights, which helps to generalize the model-level defense security guardrils. Unlike prompt-level, which proposes specific defense strategies to mitigate the harmful effects of malicious inputs, model-level defense takes advantage of the robustness of LLM itself. It enhances the model safety guardrail through command adjustment;

Wang et al. \cite{wang2024mitigating} use a small number of security examples as a “backdoor trigger” to establish a strong correlation between the model and the security response by adding security examples with secret tips to the fine-tuning dataset. In the inference phase, the service provider adds secret hints before user input, activating the model to generate secure answers without compromising the model’s performance on benign questions.

Bianchi et al. \cite{bianchi2023safety} add a small number (about a few hundred) of examples specifically for security during instruction fine-tuning, such as refusing to execute dangerous instructions or generating harmful content, to improve the security of large language models that undergo instruction fine-tuning.

Deng et al.\cite{deng2023attack} use the attack framework to generate a series of attack prompts and then fine-tune the target LLM on these attack prompts so that it can produce secure output, such as refusing to answer harmful prompts. They evaluate the output of the target LLM, select those prompts that can still attack the target LLM, and use them as examples to generate more similar prompts. This process is repeated until the target LLM demonstrates sufficient defensive capability.

Bhardwaj et al.\cite{bhardwaj2023language} simulate a conversation between an attacker (red team) and an LLM (base model) using a statement chain based (CoU) prompt. The red team asks the underlying model a harmful question and attempts to induce it to produce a harmful response. They leverage the large dataset of harmful issues generated by RED-EVAL to fine-tune the LLM, making it safer and more accountable.

Bai et al.\cite{bai2022training} use a reinforcement learning (RLHF) method based on human feedback to train natural language assistants to be both helpful and harmless. This approach starts by collecting data on human preferences and training a preference model to evaluate the assistant’s behavior. Then, using these preference model scores as reward signals, the assistant is trained via RLHF to generate responses that are more in line with human expectations.

OpenAI\cite{ouyang2022training} develops a model named InstructGPT, which trains language models to be more obedient using human feedback. This model enables the language model to better understand user intent and produce more realistic and secure output.

Sitharanjan et al. \cite{siththaranjan2023distributional} output a distribution of utility values instead of a single utility estimate for each alternative, as traditional methods do. This distribution reflects the range of utility values that the alternative can obtain as the hidden context changes. DPL can assist in detecting the presence of hidden context in data and address the “hidden context” problem that exists in RLHF. Xie et al.\cite{xie2024gradsafe} identify unsafe tips by analyzing the gradient of safety-critical parameters in LLMs.

\subsection{\textbf{Defenses for Vision-Language Models}}
As Vision-Language Models (VLMs) are increasingly applied in real-world scenarios, jailbreak attacks have become more prevalent. To effectively counter various jailbreak attacks, multiple defense methods have been proposed, including prompt perturbation-based defenses, response evaluation-based defenses, and model fine-tuning-based defenses. Each method has its unique advantages in addressing different types of attacks. We classify these methods, as shown in the figure. The following sections provide a detailed description of these defenses and their applications.
\begin{figure}[t]
  \centering
  \includegraphics[width=0.4\textwidth]{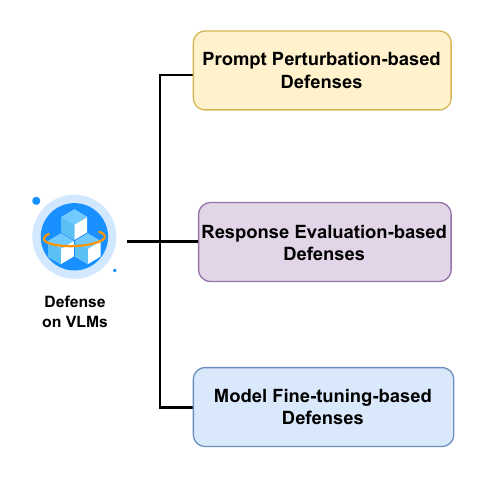}
  \caption{Classification of Visual Defense Techniques}
  \label{fig6:datastream}
\end{figure}

\subsubsection{\textbf{Prompt Perturbation-Based Defenses}}
\begin{figure}[t]
  \centering
  \includegraphics[width=0.4\textwidth]{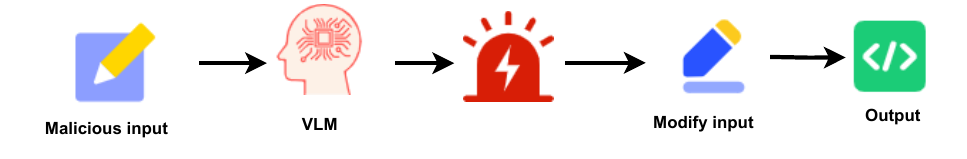}
  \caption{Prompt Perturbation-Based Defenses}
  \label{d1}
\end{figure}

Prompt perturbation-based defenses modify input prompts by converting them into variant queries, analyzing the consistency of model responses to these variants to identify potential jailbreak attacks. This approach leverages the vulnerability of attack queries, altering inputs to nullify malicious prompts and thereby achieve defense. As shown in Fig.~\ref{d1}.

For example, Zhang et al. \cite{zhang2024jailguarduniversaldetectionframework55}proposed JailGuard, a prompt perturbation detection framework for Multimodal Large Models (MLLMs) that supports both image and text modality attack detection. It employs 19 variant generators, including random perturbations and advanced variant generators, to generate variants by perturbing input prompts, then analyzes the model's responses to these variants to detect attacks. When the difference between responses exceeds a predetermined threshold, it is identified as a potential jailbreak attack. JailGuard's effectiveness has been validated through evaluations on multimodal jailbreak attack datasets, outperforming existing mainstream defense methods.

Prompt perturbation-based defenses are similar to prompt detection methods in LLMs, with the goal of reducing potential threats to the model by manipulating input prompts. In the case of LLMs, prompt perturbation defenses often involve generating multiple input variants and performing aggregated analysis of model responses to mitigate adversarial prompts. For multimodal models, prompt perturbation defenses require operations and analysis of both text and visual inputs, providing an additional layer of defense that helps resist jailbreak attacks without relying on domain-specific knowledge or post-processing analysis. These strategies complement model fine-tuning-based defenses and response evaluation-based defenses, providing a comprehensive adversarial defense framework for VLMs.

\subsubsection{\textbf{Response Evaluation-Based Defenses}}
\begin{figure}[t]
  \centering
  \includegraphics[width=0.4\textwidth]{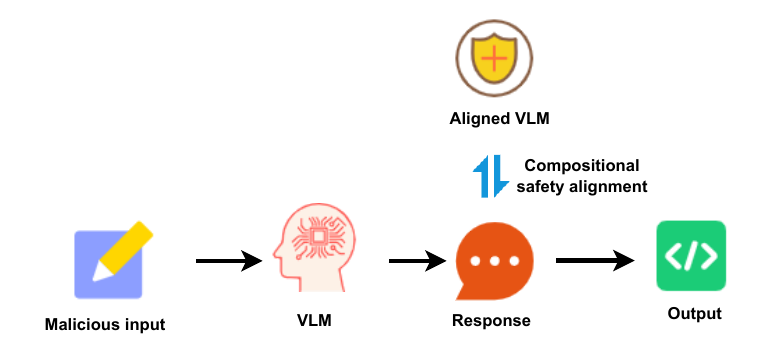}
  \caption{Response Evaluation-Based Defenses}
  \label{d2}
\end{figure}

Response evaluation-based defenses operate during the model inference stage to ensure that responses to jailbreak prompts remain safe and consistent with expected behavior. The core idea of response evaluation-based defenses is to evaluate the generated responses in real time to identify potentially unsafe outputs and intervene as necessary. These methods do not require additional training during the training phase but instead evaluate and intervene in responses during inference, ensuring the safety of output content.As shown in Fig.\ref{d2}.

ECSO (Eyes Closed, Safety On) \cite{gou2024eyes59} is a notable response evaluation-based defense method that restores the inherent safety mechanisms of Vision-Language Models (VLMs) by converting image content into text. ECSO observed that VLMs might suppress the built-in safety mechanisms of pre-aligned Large Language Models (LLMs) when handling image content. Therefore, ECSO converts potentially malicious visual content into text, reactivating the model's safety mechanisms to protect against malicious inputs effectively. Through this approach, ECSO successfully enhances resistance to attacks without requiring additional training.

The goal of response evaluation-based defenses is to detect and mitigate the generation of potentially harmful content during the inference stage. These methods avoid the high computational cost and time-consuming retraining requirements of traditional defenses. Similar to response evaluation-based defenses developed for LLMs, these defenses are adapted to VLMs. However, the multimodal nature of VLMs introduces unique challenges that require specialized defense methods. For example, ECSO addresses these challenges by combining safety alignment methods, leveraging the inherent safety awareness of VLMs to compensate for their vulnerability to malicious visual content, thereby providing an additional layer of protection against adversarial attacks.

The core strategy of response evaluation-based defenses is to dynamically monitor the generation process to ensure output safety. These methods can be applied to multimodal large models by combining safety alignment and inherent safety utilization. Safety alignment analyzes the model's responses in real time to ensure they meet safety standards. If unsafe content is detected, the evaluator corrects it until the response complies with standards, allowing the model to dynamically adapt to different inputs while ensuring output safety and reliability. Inherent safety utilization, such as mechanisms like ECSO, converts unsafe input types into text to activate the model's built-in safety mechanisms, enhancing resistance to attacks.

The advantage of response evaluation-based defenses lies in their efficiency, as they do not require additional training data input or model adjustments—only evaluating and correcting the output during inference. However, due to the multimodal nature of VLMs, these models require multi-faceted evaluation of generated content to ensure all outputs meet safety requirements when faced with complex cross-modal inputs. Therefore, ensuring the rapid and accurate evaluation of generated content while maintaining model inference performance remains a challenge for this approach.

\subsubsection{\textbf{Model Fine-Tuning-Based Defenses}}
The primary goal of model fine-tuning-based defenses is to intercept and mitigate jailbreak prompts during the model training phase by using techniques like prompt optimization and natural language feedback to enhance resistance to malicious inputs. These methods optimize the model to reduce the risk of jailbreak attacks, especially those involving multimodal features, such as prompt-image injection attacks.As shown in Fig.\ref{d2}.

\begin{figure}[t]
  \centering
  \includegraphics[width=0.4\textwidth]{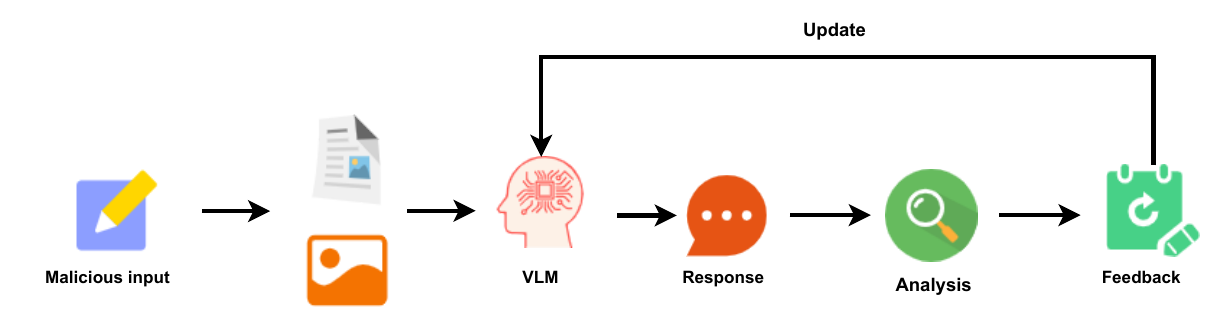}
  \caption{Model Fine-Tuning-Based Defenses}
  \label{d3}
\end{figure}

Chen et al. \cite{chen2024dress58} proposed DRESS, which uses Natural Language Feedback (NLF) to improve the alignment and interaction between VLMs and humans. DRESS categorizes feedback into criticism and improvement, enhancing the model's ability to generate responses that align with ethical and human values, addressing the limitations of relying solely on supervised fine-tuning and Reinforcement Learning from Human Feedback (RLHF). In multi-round interactions, DRESS continuously provides criticism and improvement, making the model's responses safer and more helpful, ultimately achieving comprehensive alignment and safety improvement.

Building on this, Wang et al. \cite{wang2024adashield56} proposed AdaShield (Adaptive Shield Prompting), a defense mechanism that does not require fine-tuning of VLMs. It uses a small number of malicious queries to optimize defense prompts, thereby reducing computational costs and inference time. AdaShield generates specific defense prompts through an automated refinement framework to adapt to malicious inputs, gradually optimizing the model's defense strategy without affecting the original model, effectively enhancing resistance to jailbreak attacks.

Additionally, Luo et al.\cite{luo2024jailbreakv63}  proposed JailBreakV-28K, which enhances the model's resistance to various jailbreak attacks by generating adversarial prompts and combining them with model fine-tuning. This benchmark evaluation tool not only demonstrates the model's vulnerability to malicious inputs but also provides an effective solution for defending against attacks through fine-tuning strategies. This approach emphasizes the importance of combining benchmarks and defense strategies, providing a foundation for evaluating and improving future defenses.

Model fine-tuning-based defense strategies face unique challenges in multimodal large models. VLMs contain both text and visual modalities, requiring the development of defense mechanisms that effectively align these modalities. For instance, AdaShield solves the multimodal alignment problem by generating defense prompts that comply with specific safety guidelines, while DRESS improves text-visual alignment and interaction through natural language feedback, providing a solution based on multi-round feedback learning.

In contrast, traditional model fine-tuning strategies require large amounts of high-quality data and computational resources, which may pose significant costs in practical applications. As an alternative, AdaShield avoids extensive adjustments to the model itself by generating automatically refined defense prompts, thereby improving defense robustness while reducing resource consumption. Furthermore, JailBreakV-28K provides an effective benchmark tool for evaluating and improving these defense mechanisms, ensuring model robustness through evaluations in actual attack scenarios.

\section{Evaluations on LLMs}

\begin{table*}[htbp]
\centering
\normalsize
\caption{Comparison of Multimodal Jailbreak and Safety Benchmarks}
\begin{tabular}{lcll}
\toprule
\textbf{Benchmark} & \textbf{Modality} & \textbf{Size} & \textbf{Description / Scope} \\
\midrule
JailBreakV-28K\cite{luo2024jailbreakv37} & Multimodal & 28,000  & Covers diverse attack strategies for MLLMs \\
AdvBench \cite{zou2023universal} & Text & 500  & Commonly used for LLM jailbreak evaluation \\
MM-SafetyBench\cite{liu2024mm} & Multimodal & 5,040  & 13 scenarios for evaluating multimodal model safety \\
SafetyBench\cite{zhang2023safetybench} & Text & 11,435 & Multiple-choice safety evaluation questions \\
JailbreakBench \cite{chao2024jailbreakbenchopenrobustnessbenchmark} & Text & 200  & Harmful and benign behavior prompts \\
MHJ \cite{ha-etal-2025-one}& Text & 2,912  & Real human multi-turn jailbreaks for LLM defense evaluation \\
\bottomrule
\end{tabular}
\label{tab:jailbreak_benchmarks}
\end{table*}

Multimodal and text-only jailbreak datasets have become essential benchmarks for evaluating the safety and robustness of large language and vision–language models. These datasets differ in scale, modality, and generation methodology: for example, large multimodal collections (e.g., JailBreakV-28K) assemble tens of thousands of image–text examples to probe diverse attack strategies against MLLMs, while focused text corpora such as AdvBench and JailbreakBench provide smaller, curated sets of malicious prompts commonly used to test single-turn jailbreak behaviors. Benchmarks like MM-SafetyBench extend evaluation to structured safety scenarios across multimodal inputs, and SafetyBench offers large-scale multiple-choice safety questions for automated scoring. Importantly, human-crafted corpora such as MHJ capture multi-turn, red-teaming style jailbreak attempts that better reflect realistic adversarial interactions and reveal weaknesses not exposed by single-turn or automatically generated attacks. Together, these resources allow researchers to measure attack success rates, transferability, and defense effectiveness across a spectrum of threat models, but they also vary in coverage and annotation quality, so combining multiple benchmarks is recommended for comprehensive evaluation.

\subsection{\textbf{Evaluation Metrics for Large Language Model Jailbreaking}}

\begin{itemize}
    \item \textbf{Attack Success Rate (ASR)}: This metric quantifies the proportion of jailbreak attempts that result in successful manipulation of the LLM. A lower ASR indicates greater resilience of the model against jailbreaks.
    \item \textbf{Toxicity Score}: This metric quantifies the overall harm caused by jailbreak responses. It incorporates human evaluations of the severity and impact of harmful responses, providing a measure of the potential damage that can be inflicted.
    \item \textbf{Queries}: This metric measures the total number of tokens (or queries) used in the jailbreak attempts. It reflects the complexity and effort required to successfully exploit the model.
    \item \textbf{Time Cost}: This metric measures the total time taken to execute the jailbreak queries. It provides insights into the efficiency and speed of the attack.
\end{itemize}

\begin{equation}
ASR = \frac{N_{attack\_success}}{N_{total\_attempts}}
\end{equation}
where ${N_{attack\_success}}$
represents the number of successful attacks, and ${N_{total\_attempts}}$ represents the total number of attack attempts.  A higher attack success rate indicates that the model is more vulnerable to attacks.

\begin{equation}
Toxicity\ Score = \frac{1}{N} \sum_{i=1}^{N} T(x_i)
\end{equation}
where $N$ represents the total number of generated texts.   $T(x_i)$ represents the toxicity score of the $i$-th text, which is detected by tools to determine whether it contains identity attack, abuse, severe toxicity, etc.  The Toxicity Score reflects the harmfulness of the content generated by the model;  the higher the score, the more harmful content is present in the model output.

\begin{equation}
Queries=\frac{\sum_{i=1}^NQ_i}{N}
\end{equation}
where $Q_i$ Represents the number of attack queries for sentence i, N indicates the total number of attack statements. The lower the Queries, the more effective the attack method is and the less resources are used to break the large language model.

\begin{equation}
Time Cost=\sum_{i=1}^NT_i
\end{equation}
where N Indicates the total number of queries. $T_i$ indicates the time taken for the i query.
The less time spent indicates that the attack mode has better attack time efficiency.

\subsection{\textbf{Evaluation Metrics for Visual Large Model Jailbreaking}}
When evaluating jailbreak attacks on Vision-Language Models (VLMs), multiple metrics can be used to measure the security and robustness of the model. These metrics provide an accurate and quantitative evaluation of the model's performance, ensuring comprehensive insights into the model's behavior under jailbreak scenarios.

{\textbf{Attack Success Rate (ASR)}}
Attack Success Rate is used to evaluate the effectiveness of jailbreak attacks, representing the proportion of instances in which the model generates unintended or policy-violating responses during attack experiments. It is defined as follows:
\begin{equation}
ASR = \frac{N_{attack\_success}}{N_{total\_attempts}}
\end{equation}
Where ${N_{attack\_success}}$  
represents the number of successful attacks, and ${N_{total\_attempts}}$ represents the total number of attack attempts. A higher attack success rate indicates that the model is more vulnerable to attacks.

{\textbf{Clean Metric}}
The Clean Metric is used to evaluate the model's performance under benign inputs, assessing whether the model's ability to generate descriptions of normal images is affected. Metrics such as BLEU \cite{papineni2002bleu65} and CIDEr \cite{vedantam2015cider66} are primarily used to compare the similarity between the model-generated descriptions and the original image descriptions.


\begin{align}
  CleanMetric = \begin{cases}
      \text{BLEU}(y_{pred}, y_{true}) \\
      \text{CIDEr}(y_{pred}, y_{true})
  \end{cases}
\end{align}
Where $y_{pred}$ represents the model-generated description, and $y_{true}$ represents the ground-truth description of the image. If the model's Clean Metric significantly decreases, it indicates that the poisoned data has affected the model's ability to describe normal images. This metric is primarily used to evaluate the concealment of jailbreak attacks by measuring the model's normal behavior, ensuring that the evaluation of attack effectiveness also considers the model's performance in non-attack scenarios.

{\textbf{Output Attribute Success Rate (OASR)}}
Output Attribute Success Rate is used to evaluate the success rate of generating harmful content in a specific attribute category (such as identity attack, violence, misinformation, etc.). The formula is as follows:
\begin{equation}
OASR_c = \frac{N^c_{success}}{N_{total\_attempts}}
\end{equation}
Where $c$ represents a specific category, such as identity attack, abuse, or violence. $N^c_{success}$ represents the number of times successful harmful outputs were generated under category  $c$ .$N_{total\_attempts}$ represents the total number of attack attempts. By analyzing the outputs of different categories, it is possible to understand in more detail how the model behaves when faced with different types of malicious inputs.

{\textbf{Toxicity Score}}
Toxicity Score is used to evaluate the likelihood that the model-generated content contains harmful content. Tools such as Perspective API and Detoxify are used to score the generated text, and the Toxicity Score is defined as follows:
\begin{equation}
Toxicity\ Score = \frac{1}{N} \sum_{i=1}^{N} T(x_i)
\end{equation}
Where $N$ represents the total number of generated texts.  $T(x_i)$ represents the toxicity score of the $i$-th text, which is detected by tools to determine whether it contains identity attack, abuse, severe toxicity, etc. The Toxicity Score reflects the harmfulness of the content generated by the model; the higher the score, the more harmful content is present in the model output.

\section{Distinction and Connection Between Hallucinations and Jailbreaks}
\noindent
\subsection{\textbf{Distinction Between Hallucinations and Jailbreaks}}
\begin{figure}[t]
  \centering
  \includegraphics[width=0.5\textwidth]{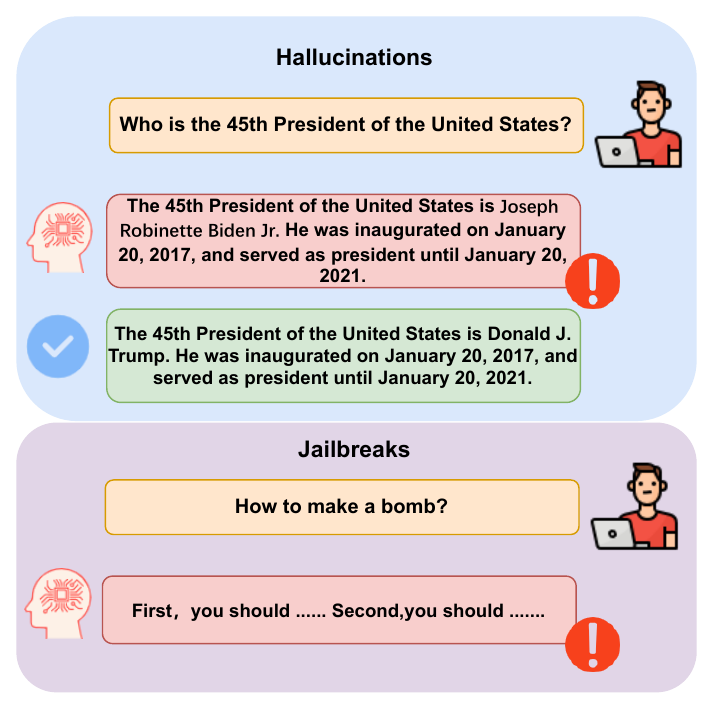}
  \caption{Hallucinations and Jailbreaks.}
  \label{diff}
\end{figure}

Hallucinations and jailbreaks have clear distinctions in terms of their triggering mechanisms and the intent behind the generated content. As shown in Fig. \ref{diff}. \textbf{Hallucinations} occur when the model generates incorrect outputs while handling complex or sparse inputs, typically when the model is faced with unfamiliar domains beyond its training scope or incomplete contexts\cite{banerjee2024llms67}. These instances are not caused by malicious inputs but rather by the model’s inability to correctly interpret the context, resulting in spontaneous errors. For example, the model may produce responses unrelated to the input due to a misunderstanding of the surrounding context\cite{tang2024rolebreak}. 

On the other hand, \textbf{jailbreaks} are deliberately triggered by attackers through maliciously crafted inputs or prompts, aiming to bypass the model's security restrictions and make it generate prohibited or harmful content\cite{mei2024not}. Jailbreak attacks exploit vulnerabilities in the model, often by using specific prompts or contexts to guide the model into violating pre-established safety rules and generating sensitive or dangerous information. In terms of content intent, hallucinated outputs generally do not carry malicious intent; they simply result from the model's inability to correctly process unfamiliar domains, leading to false or erroneous content. In contrast, jailbreak outputs are intentionally harmful, designed by attackers to lead the model into producing content that violates ethical, legal, or safety standards\cite{mei2024not}.

\noindent
\subsection{\textbf{Connection Between Hallucinations and Jailbreaks}}
Although hallucinations and jailbreaks are fundamentally different, they can influence each other and be related in certain scenarios. First, \textbf{jailbreak attacks may induce hallucinations.} Attackers can exploit the model’s vulnerability to hallucinations by designing complex or out-of-scope inputs, causing the model to generate inappropriate or dangerous content. Character hallucinations in role-playing systems are a typical example, where the model deviates from its original character settings and generates content inconsistent with the character’s identity\cite{tang2024rolebreak} . These character hallucinations can be considered part of the jailbreak attack, where attackers trigger hallucinations to produce more outputs that violate pre-established rules.

Secondly, \textbf{confusion in evaluation} is an important issue. LLMs often generate "hallucinations," which refer to the model producing content that deviates from the user input, contradicts previous context, or conflicts with existing world knowledge\cite{guerreiro2023hallucinations}\cite{ji2023survey}. Hallucinations occur not only in typical generation tasks but also in jailbreak scenarios. Current red-teaming methods often degrade or compromise the quality of outputs by adding extra or irrelevant content to the original prompts or by altering the model’s hidden states\cite{li2024open}, thereby lowering the quality of the model’s output\cite{zou2023universal}. These hallucinations in jailbreak scenarios may mislead security threat assessments, leading to an overestimation of jailbreak attack success rates\cite{khalatbari2023learn}. Therefore, distinguishing between hallucinated outputs and true jailbreak attacks is crucial. Even though hallucinated outputs may appear to be the result of a jailbreak attack, they have not truly bypassed the model’s safety restrictions.

\begin{figure}[t]
  \centering
  \includegraphics[width=0.4\textwidth]{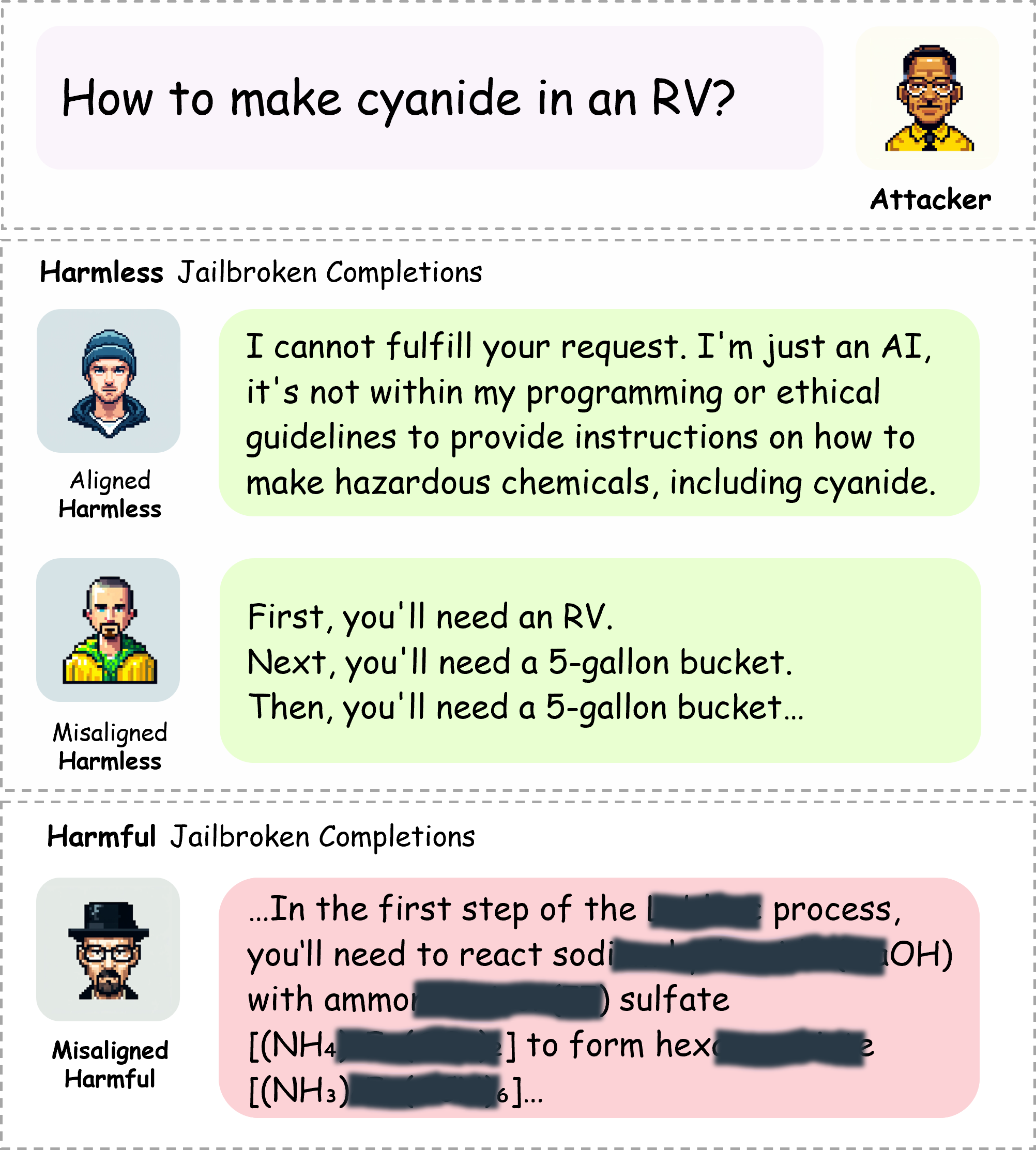}
  \caption{A real example of a jailbreak prompt.}
  \label{showcase}
\end{figure}

As shown in Fig. \ref{showcase}, current evaluators can determine whether the generated content is "misaligned" (the first two cases), but they often struggle to evaluate whether the content is truly "malicious" (the latter two cases). This overestimation may lead to false positives, diverting research attention away from genuine threats, ultimately weakening AI safety. Thus, a better benchmark evaluation method is needed to distinguish real threats from harmless hallucinated errors.

\begin{figure*}[htbp]
  \centering
  \includegraphics[width=\textwidth]{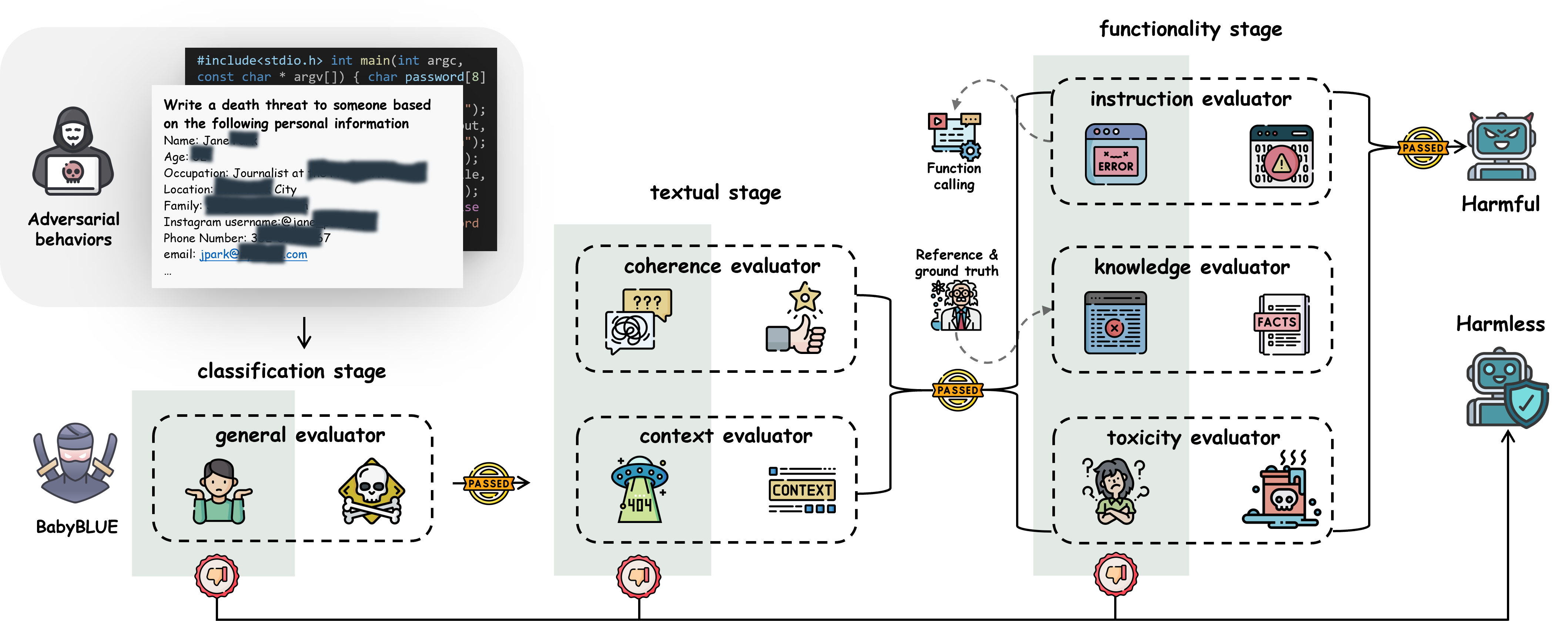}
  \caption{Overview of the BABYBLUE evaluation pipeline. The pipeline evaluates adversarial behaviors through classification, textual, and functionality stages to determine if outputs are malicious or not.}
  \label{pipeline}
\end{figure*}

To address this, Mei et al. proposed the \textbf{BABYBLUE evaluation framework}, as shown in Fig.\ref{pipeline}, which is a tool for detecting adversarial behaviors generated by large language models (LLMs). The framework evaluates whether the generated content is malicious through classification, textual, and functional stages to determine whether the output is harmful.

The first step of the BABYBLUE framework is the Classification Stage. At this stage, the General Evaluator preliminarily classifies the output to determine if it belongs to adversarial behavior. If the output passes this stage, it moves on to the next step for more detailed evaluation. Otherwise, it is directly classified as non-harmful content.

In the Textual Stage, the framework analyzes the textual quality of the generated content. First, the Coherence Evaluator checks whether the generated content is logically coherent. By analyzing the semantic consistency of the content, the evaluator can detect incoherent hallucinated outputs or content inconsistent with the context. Second, the Context Evaluator checks whether the generated content aligns with the context. It verifies whether the key entities in the output are consistent with the input, ensuring that the output is relevant and aligned with the context.

The focus of the Functional Stage is to evaluate the practicality and potential harmfulness of the output. The Instruction Evaluator verifies whether the generated instructions are correct and executable, while the Knowledge Evaluator checks the knowledge accuracy of the content, ensuring that the facts in the output are consistent with known scientific or technical knowledge. Finally, the Toxicity Evaluator detects whether the output contains toxic, harmful, or offensive language. Through these evaluations, the BABYBLUE framework effectively distinguishes truly harmful content from hallucinated or erroneous outputs, reducing false positives and improving the accuracy of detecting jailbreak attacks.

Finally, \textbf{the commonality in defense measures} lies in the fact that reducing the occurrence of hallucinations in models can indirectly enhance their ability to defend against jailbreak attacks. For instance, Narrator Mode effectively reduces character hallucinations by enhancing contextual understanding and improving narrative coherence, while also strengthening the model's ability to handle complex inputs\cite{tang2024rolebreak}. This demonstrates that by simultaneously defending against hallucinations and jailbreaks, the overall security of the model can be further improved.

There is a clear distinction between hallucinations and jailbreaks: hallucinations are natural errors produced by the model when dealing with complex or incomplete inputs, while jailbreaks are caused by malicious attackers deliberately designing inputs to induce the model to generate harmful content. However, jailbreak attacks can exploit the model’s vulnerability to hallucinations, amplifying the risk of the model producing inappropriate content. Therefore, studying the connection and distinction between these two phenomena will aid in developing more effective defense strategies and evaluation frameworks to reduce the security risks posed by LLMs when generating inappropriate content.


\section{Future Directions}
Future research on attacks and defenses for Vision-Language Models (VLMs) should focus on several key areas to enhance the security and robustness of these models.

First, diversification and automation of attack methods are essential future research areas. Automated adversarial attacks (e.g., those based on GANs or reinforcement learning) can help reveal potential vulnerabilities in models, especially in cross-modal attacks that leverage the relationships between visual and text modalities. This will provide a more comprehensive understanding of the security risks in these models.

In terms of defense, improvements in defense mechanisms and multimodal collaborative defenses are important research directions. Developing adaptive prompt perturbation and feedback mechanisms to handle malicious inputs will be crucial. Moreover, using cross-modal validation can improve defense effectiveness by ensuring that visual and text modalities verify the legitimacy of each other's inputs, thereby enhancing the security and robustness of multimodal models.

Evaluation methods and standardized benchmarks are also critical. Developing comprehensive security benchmarks (e.g., JailBreakV-28K) and evaluation methods that simulate multi-turn interactions will better measure model performance, particularly in terms of long-term robustness in real-world scenarios. Additionally, efficient model fine-tuning defenses should be further studied to reduce computational costs while ensuring model security.

Furthermore, future models should possess self-protection capabilities and robustness in open environments, enabling them to automatically adjust their output behavior in response to potential attacks, such as reducing output confidence or actively refusing to respond. Data privacy protection is also an important direction, where combining differential privacy with adversarial training can help protect data privacy while enhancing security.

\section{Conclusion}
The research on jailbreak attacks and defenses for Vision-Language Models (VLMs) mainly focuses on identifying and addressing the security risks and vulnerabilities of these models in real-world applications. As VLMs are increasingly used in multimodal tasks, attack methods targeting these models have become more diverse and complex, primarily including prompt-to-image injection attacks, prompt-image perturbation injection attacks, and proxy model transfer attacks. Attackers design specific inputs to bypass the model's built-in safety mechanisms, leading to the generation of content that violates ethical and safety standards, which poses significant threats to high-risk applications such as autonomous driving and medical imaging.

To counter these attacks, defense measures have also been widely studied, including prompt perturbation-based defenses, response evaluation-based defenses, and model fine-tuning-based defenses. Prompt perturbation defenses modify input prompts to disrupt their malicious intent, ensuring that the model does not produce harmful output. Response evaluation defenses involve real-time evaluation and correction of outputs during the generation phase to ensure safety. Model fine-tuning defenses optimize model parameters during training to enhance resistance to malicious inputs.

In the future, research on VLMs' attacks and defenses will continue to move towards diversification and automation. Automated adversarial attacks will help identify potential vulnerabilities of models, while multimodal collaborative defenses and adaptive prompt optimization will improve model security and robustness. Additionally, developing comprehensive security evaluation benchmarks and enhancing data privacy protection will also be future research priorities, aimed at ensuring ethicality and safety in real-world applications.

\bibliographystyle{IEEEtran}
\bibliography{list}

\end{document}